\documentclass[sigconf]{acmart}

\AtBeginDocument{%
  }

\usepackage[english]{babel}
\usepackage{blindtext}
\usepackage{totpages}
\usepackage{tikz}
\usepackage{amsmath}
\usepackage{enumitem}
\usepackage{xcolor}
\usepackage{subcaption}
\usepackage{xspace}
\usepackage{booktabs}
\usepackage{threeparttable}
\usepackage{pifont}
\usepackage{bbding}

\newcommand{\paraskip}{\vspace{2pt}}
\newcommand{\parahead}[1]{\paraskip\noindent\textbf{#1}}
\newcommand{\chatgpt}{\textsf{ChatGPT}\xspace}
\newcommand{\gemini}{\textsf{Gemini}\xspace}
\newcommand{\grok}{\textsf{Grok}\xspace}
\newcommand{\doubao}{\textsf{Doubao}\xspace}
\newcommand{\yuanbao}{\textsf{Yuanbao}\xspace}
\newcommand{\qwen}{\textsf{Qwen}\xspace}

\renewcommand\footnotetextcopyrightpermission[1]{}

\author{
{\rm Jiayang Xu, Xiangjie Huang, Zijie Li, Antariksh Verma, Zili Meng}\\
Hong Kong University of Science and Technology 
}

\setlist[itemize]{leftmargin=*}
\setlist[enumerate]{leftmargin=*}

\setitemize{leftmargin=4mm}
\setenumerate{leftmargin=4mm}
\setlist{noitemsep,topsep=0pt,parsep=0pt,partopsep=0pt}

\begin{document}

\title{Make a Video Call with LLM:\\A Measurement Campaign over Six Mainstream Apps}

\begin{abstract}
  In 2025, Large Language Model (LLM) services have launched a new feature -- AI video chat -- allowing users to interact with AI agents via real-time video communication (RTC), just like chatting with real people. 
  Despite its significance, no systematic study has characterized the performance of existing AI video chat systems. 
  To address this gap, this paper proposes a comprehensive benchmark across four dimensions: quality, latency, internal mechanisms, and system overhead. 
  Using custom testbeds, we further evaluate six mainstream AI video chatbots with this benchmark. 
  We also build an online platform for user study.
  The measurement leads to interesting findings that could be beneficial to the future optimizations.
  For example, the network latency of AI video chat matters not as much as human video chat.
  The capabilities of AI agents matters most in the user experience.
  % This work provides the research community a baseline of real-world performance and identifies unique system bottlenecks. 
  Our benchmarking results also open up several research questions for future optimizations of AI video chatbots.\\
  \textbf{Availability:} \url{https://callarena.net/} for the online evaluation platform and our open-sourced dataset and testbed.
\end{abstract}

\maketitle
\pagestyle{plain}

%-------------------------------------------------------------------------------
\section{Introduction}
%-------------------------------------------------------------------------------

In 2025, mainstream Large Language Model (LLM) services (e.g., \chatgpt~\cite{chatgpt-official}, \gemini~\cite{gemini-official}, and \grok~\cite{grok-official}) started to offer a new feature -- AI video chat -- in which users converse with an AI agent over real-time video communication (RTC) much as they would with another person.
By letting the user continuously share visual context with the agent, AI video chat addresses the limitations of text- or audio-only interactions and opens up scenarios in which the AI can perceive and reason about the world around the user.
This direction is reinforced by the next generation of consumer hardware: AI-powered smart glasses (e.g., Ray-Ban Meta with Live AI~\cite{meta_rayban_live_ai}) are being designed around always-on, hands-free video understanding, making AI video chat the primary interaction modality rather than one feature among many -- the user's hands and eyes are occupied with the physical world, and a continuous video feed is the most natural input channel.
Economically, the traditional video chat industry is already a mature market valued in the tens of billions of dollars~\cite{fortune_vc}, while global spending on LLMs is projected to approach \$1.3T by 2032~\cite{bloomberg_bi_genai}.
The integration of AI with video chat therefore signals substantial business opportunities and societal impact.

Despite its significance, there has not been any systematic study characterizing the performance of existing AI video chat systems. 
Although the status-quo network stack might be similar, AI video chat applications fundamentally differ from existing applications such as traditional video chat, AI text/audio chat, and video analytics. 
The difference comes from multiple dimensions -- latency and quality metrics, and the new scenarios that are enabled by the AI video chat.
For example, compared to traditional video chat, the receiver of the video changes from human to machine.
% where latency is dominated by client- or network-side delays, AI video chat introduces LLM inference into the loop, shifting the primary performance bottleneck to computational processing time. 
This also alters evaluation priorities: rather than focusing solely on perceptual quality and end-to-end latency, the emphasis shifts toward the accuracy and responsiveness of the AI’s output. 
Compared to text- or audio-only AI systems where the processing can be stateless and all the conversation contexts are sent each time, AI video chat must handle continuous high-bitrate video streams, which significantly challenge the computational demands and transmission patterns. 
% Moreover, AI video chat requires maintaining visual memory during runtime, a capability unnecessary in text- or audio-only chat. 
Finally, unlike video analytics, AI video chat is fundamentally interactive, demanding a conversational flow. 

% In other words, the fundamental architectural and performance differences of AI video chat render prior evaluation frameworks inapplicable, motivating the need for a specialized suite of metrics tailored to its unique characteristics. 

\begin{table}[t]
    \centering
    \begin{tabular}{l|ccc}
        % \toprule
        \hline
        \  & Release & Support Video Chat & Company \\ 
        % \midrule
        \hline
        \chatgpt & Nov 2022 & Dec 2024 & OpenAI \\ 
        % \hline
        \gemini & Feb 2024 & Mar 2025 & Google \\ 
        % \hline
        \grok & Nov 2023 & Apr 2025 & xAI \\ 
        % \hline
        \doubao & Aug 2023 & May 2025 & ByteDance \\ 
        % \hline
        \yuanbao & May 2024 & Jun 2025 & Tencent \\ 
        % \hline
        \qwen & Nov 2025 & Nov 2025 & Alibaba \\ 
        % \bottomrule
        \hline
    \end{tabular}
    \caption{Overview of tested applications}
    \label{tab:apps-overview}
\end{table}

To address this gap, in this paper, we propose a benchmark that can comprehensively evaluate the performance of AI video chat, with carefully designed evaluation metrics spanning four dimensions -- quality, latency, internal mechanisms, and system overhead. 
On the dimension of quality, we analyze several unique use cases that emerge from the usage of AI video chat -- e.g., answering a question based on the video minutes ago. 
For latency, we investigate both the time required to set up a video chat and the response delay of the AI agent.
Regarding internal streaming mechanisms, we study how the design choices of different protocols, bitrate, frame rate, and so on are reflected. 
We further conduct a series of microbenchmarks to measure system overhead. 
These studies, to the best of our knowledge, evaluate the performance of AI video chat comprehensively for the first time. 
% To evaluate response quality in AI video chat scenarios, we further construct a dataset derived from three representative aspects widely discussed on the Internet. 
% We test AI agents on this dataset in conjunction with a subset of an existing open-source benchmark that measures general streaming video understanding ability.

We further evaluate the performance of six mainstream AI video chatbots using our benchmark, as detailed in Table~\ref{tab:apps-overview}. 
These applications are selected based on their popularity and regional diversity: \chatgpt (OpenAI), \gemini (Google), \grok (xAI), \doubao (ByteDance)~\cite{doubao-official}, \yuanbao (Tencent)~\cite{yuanbao-official} and \qwen (Alibaba)~\cite{qwen-official}. 
The first three chatbots are the only video-enabled applications from the global top 15~\cite{top_ai_chatbots}, while the last three are the only ones with this feature among the top five in China~\cite{top_ai_apps_china_2026}. 
We developed testbeds based on both real phones and cloud-emulated phones to evaluate performance both faithfully and broadly.
The measurements were conducted across four countries and five regions in two rounds -- August 2025 and March 2026 -- with each round running continuously for at least a week.
In total, the two rounds accumulated more than 120 hours of active video-chat session time.
Leveraging these testbeds, we evaluate each application on its supported platforms (cloud or local) and use the cloud infrastructure to measure across multiple geographic regions.

Beyond the objective benchmark, we deployed an online video-chat battle platform to gather subjective user feedback.
In each session, a user concurrently runs a video chat with two randomly selected applications, asks both the same prompts, and votes for the one that delivered the better experience.
To date, the platform has logged over 200 head-to-head battles, with every application appearing in more than 80 of them, providing the first systematic user-rated comparison of AI video chat services.

We have presented several major findings based on the latest collected results as below. The detailed results with more comprehensive findings come in \S\ref{sec:results} and \S\ref{sec:user-study}. We have released the dataset and cloud testbed associated with this paper (see the link in the abstract) to facilitate future research in the community.

\begin{figure}[t]
    \centering
    \includegraphics[width=1\linewidth]{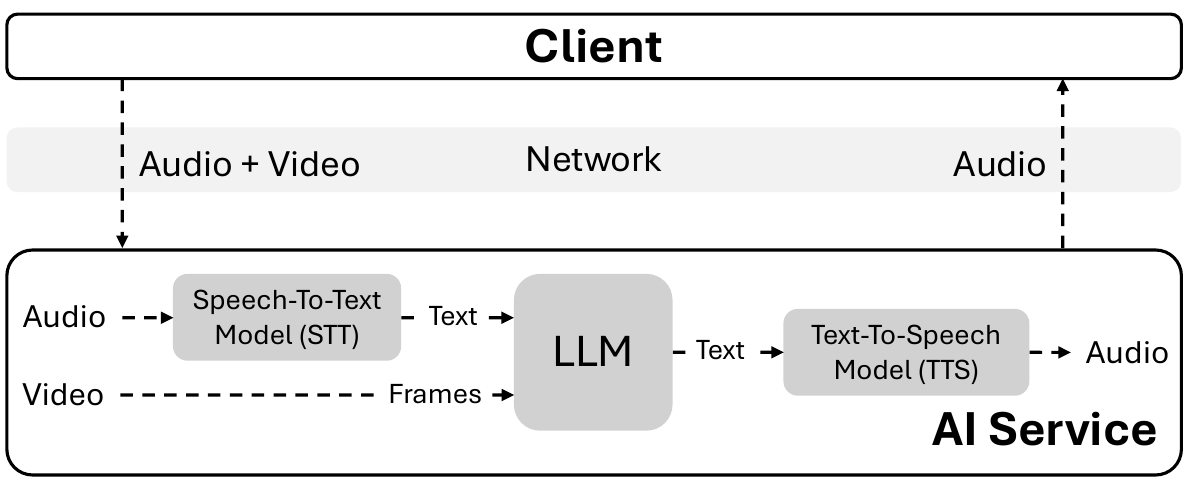}
    \caption{AI video chat paradigm}
    \label{fig:two-types}
\end{figure}

\parahead{Finding-1: Response delay is far from seamless conversation, and the causes are heterogeneous.}
Every chatbot we measured averages more than 2.8\,s of response delay (from \yuanbao at 2.85\,s up to \qwen at 4.44\,s), well above the sub-second threshold for natural human conversation~\cite{itu-t_g114_2003}.
The heavier-tailed three -- \gemini, \chatgpt, and \doubao -- exceed 9\,s at p99.
The causes are mixed, ranging from the inference time of the AI agent itself to scheduling-induced waiting at peak hours and to cross-region network round-trips.

\parahead{Finding-2: The underlying network stack has not converged.}
The design choices of the underlying network stack still differ widely across applications.
Five of the six chatbots (\chatgpt, \grok, \doubao, \yuanbao, \qwen) carry media over RTP/RTCP, while \gemini uses QUIC end-to-end; the per-app uplink video bitrate spans roughly $4\times$ ($\sim$0.5\,Mbps for \gemini up to $\sim$1.9\,Mbps for \chatgpt and \grok) and the frame rate spans $30\times$ ($\sim$1\,fps for \gemini and \grok up to $\sim$30\,fps for \chatgpt and \qwen).
This degree of dispersion is unusual for a class of mature RTC applications and indicates that the engineering trade-offs around streaming AI inputs are still being explored.

\parahead{Finding-3: Reducing network delay yields little gain in user experience.}
On our online arena, randomly injecting up to 2\,s of additional delay on one side does not measurably change which side users prefer in a blind A/B comparison; the gap between apps in the user-perceived ranking is dominated by which app the user is talking to, not by how fast it replies.
Purely network-side latency optimizations (e.g., shaving RTT) are therefore unlikely to move the user-experience needle on their own; the higher-payoff levers are AI-side processing time and response content quality.

\parahead{Finding-4: AI-side capability gaps remain wide on use cases unique to AI video chat.}
Even setting latency aside, the agents differ dramatically on capabilities that only show up in the video setting.
On \emph{visual content memory}, \chatgpt can answer questions about a video frame from 10+ minutes ago, while \yuanbao and \qwen retain nothing once the relevant frame leaves the screen.
On \emph{proactive output}, the gap is even sharper: only \doubao occasionally fires output triggered by a visual event (within 10\,s of the trigger on $\sim$80\% of trials), while the other five remain passive and respond only after the user speaks.

% \parahead{Finding-5.}
% \chatgpt, \doubao, and \yuanbao video encoding bitrate are adaptive to content complexity. 
% However, \gemini shows little difference. It encoding at a low bitrate \textasciitilde 500 Kbps for high/low motion videos. 
% On the other hand, \grok is not addaptive to video content. 
% It maintains a relatively constant high bitrate at \textasciitilde 2 Mbps all the time.  

% \parahead{Finding-6.}
% All tested applications except \chatgpt frame rate on the network is much lower than traditional video chat. 
% Even though \chatgpt has 20\textasciitilde30 fps, we think that the actual frequency as well as number of frame processed by cloud will be much less considering the AI inference time restriction. 

% \parahead{Finding-7.}
% How uplink bandwidth constraints affect response quality. 

% (there are also results on video setup time, ai response duration\&speaking rate, video send burst/pacing, network protocols, input modality)

% data rate: adaptive video bitrate streaming

% frame rate: much lower than traditional RTC

% protocol:

% MLLM's input modalities: audio to text vs. raw audio

% response delay: > 1.5 seconds, affected by scheduling (daytime and nighttime)

% response quality: general ability, rtc-specific ability (visual named entity recognition, math problem solving, memory ability)
% impact of bandwidth limitation: uplink (whether video stream can be successfully transmitted), downlink (response audio quality)

%-------------------------------------------------------------------------------
\section{Background}
\label{sec:background}
%-------------------------------------------------------------------------------

\parahead{Measurements and Benchmarks over RTC applications.}
% Video conferencing applications, such as Zoom, Google Meet, and Microsoft Teams, have gained prominence since the COVID-19 pandemic.
Numerous measurement studies were conducted in the recent years to understand the RTC performance of different applications, including video conferencing~\cite{imc21measurement,imc2022measurementzoom,bentaleb2025evaluation,imc2021canyouseemenow}, Augmented Reality/Virtual Reality (AR/VR)~\cite{cheng2022reality}, and cloud gaming~\cite{imcxiaokun}.
In these studies, a wide range of Quality of Service (QoS) metrics -- including latency, data rate, and image quality -- were recorded and analyzed.
% Similarly, several studies have measured the performance of Augmented Reality/Virtual Reality (AR/VR) applications~\cite{cheng2022reality}.
% These AR/VR systems are also built on RTC technologies, yet they impose even more stringent requirements on image quality and latency. 
In this paper, we investigate a new category of RTC applications: those that serve AI agents. 
We will explain why it is necessary to measure AI video chats using a different way in \S\ref{sec:motivation}. 

% The key distinction lies in the communication paradigm, which has shifted from human-to-human to human-to-agent.
% This shift has led to significant changes in both QoS and Quality of Experience (QoE) metrics, thereby motivating the measurement study presented in this work.

\parahead{AI video chat paradigms.}
To measure and benchmark the AI video chat applications, we first need to understand how they work (see Figure~\ref{fig:two-types}).
An AI video chat system typically streams audio and video from a user's device to an LLM hosted on cloud computing infrastructure. 
The system, in turn, generates and returns a synthesized audio response. 
It often employs a cascaded pipeline: a speech-to-text (STT) module first transcribes user audio; an LLM then processes this text alongside the video frames; and finally, a text-to-speech (TTS) module converts the model's text response into audio. 
Meanwhile, some systems directly take the audio as the input and output without STT and TTS~\cite{zeng2024glm,defossez2024moshi}. 
% The first paradigm takes an end-to-end approach, enabling the LLM to directly process raw audio and video streams to generate an audio output. 

% \parahead{Overview of tested applications.}
% We conducted an evaluation of five popular AI applications from leading technology firms, summarized in Table~\ref{tab:apps-overview}. 
% The recent and rapid rollout of AI video chat — all within the last year and most within six months — underscores its status as a significant emerging feature with strong backing from the industry. 
% With the exception of \chatgpt, which restricts video chat to its paid tiers, all other applications provide this feature for free. 
% Our testing was performed on the paid tier for \chatgpt and on the free tiers for the remaining applications. 

%-------------------------------------------------------------------------------
\section{Motivation}
\label{sec:motivation}
%-------------------------------------------------------------------------------
% AI video chat applications are at the intersection between the traditional video chat applications (e.g., Zoom and WhatsApp) and the AI text/audio chatbots (e.g., GPT-4o). 
\begin{figure}[t]
    \centering
    \includegraphics[width=0.7\linewidth]{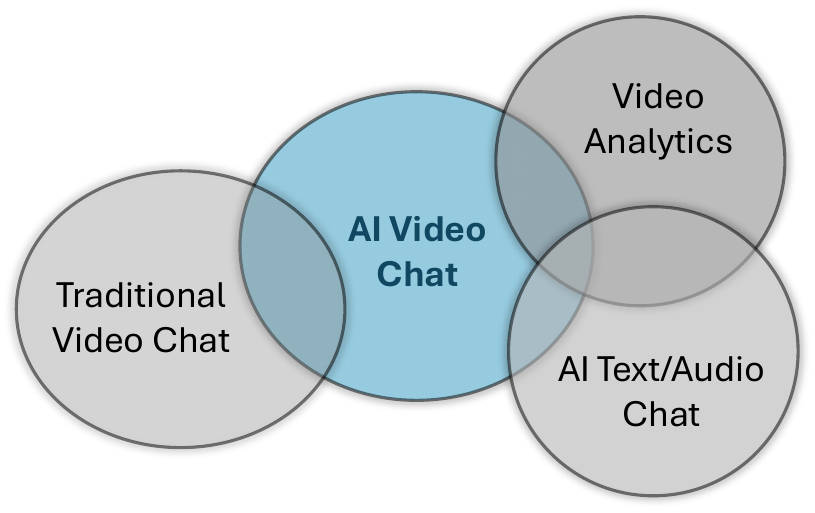}
    \caption{AI video chat differs from related applications}
    \label{fig:venn}
\end{figure}

AI video chat applications share some similarities with applications such as traditional RTC, traditional AI chat, and video analytics.
Nevertheless, we identify that AI video chat has a series of non-trivial outstanding differences that motivate us to redesign the benchmark for testing. 
Figure~\ref{fig:venn} illustrates the relationships between these applications using a Venn diagram. 
% Yet, they are different from these two categories in a non-trivial way. 

\parahead{AI video chat vs. Traditional video chat.}
The key distinctions between AI and traditional video chat are summarized below:
\begin{itemize}
    \item Receivers shift from human to agent. The video stream required for AI agents is different from human beings. This result in a new set of quality of experience metrics. This can also be reflected by the streaming parameters -- e.g., the frame rate can be much lower for agents. 
    \item Usage asks for new capabilities. For example, when using AI video chatbot, the question might be related to the video minutes ago, while in traditional RTC only the current video frame matters.
    \item Non-network bottlenecks attract more attention. In traditional video chat, delays arise mainly from network transmission and client-side processing. In AI video chat, however, the primary bottleneck is server-side, dominated by the heavy autoregressive inference of the LLM. 
\end{itemize}
% i) The performance bottlenecks are different. 

% ii) The evaluation objectives are redefined. 
% Traditional video chat aims to optimize for human perceptual quality, using metrics like the Structural Similarity Index Measure (SSIM)~\cite{ssim-understand} or Video Multimethod Assessment Fusion (VMAF)~\cite{vmaf-official}. 
% For AI video chat, however, these perceptual metrics are largely irrelevant. 
% The objective shifts to the accuracy and latency of the model's response. 

\parahead{AI video chat vs. AI text/audio chat.}
The primary feature distinguishing AI video chat from text- or audio-only chat is the addition of a continuous video input stream. 
This multimodal input introduces two significant changes: 
\begin{itemize}
    \item Increased latency. The constant processing of video data adds a substantial computational burden to the AI's backend. This heightened inference load results in longer processing times, leading to higher latency compared to interactions that only involve text or audio. 
    \item Stateful processing of video. Unlike AI text/audio chat, video chat demands a more advanced, multimodal memory. It need to perform stateful processing, continuously integrating visual information from the video with the ongoing dialogue -- a unique challenge not found in text or audio-only formats.
\end{itemize}
% i) 
% ii)  

\parahead{AI video chat vs. Video analytics.}
While AI video chat also shares traits with video analytics -- both use AI to process video input -- their fundamental purpose differs. 
AI video chat is built for interactive, real-time conversation, whereas video analytics focuses on passive observation.  

Therefore, differences either in performance, architecture, or evaluation objectives distinguish AI video chat from related technologies. 
This distinction highlights a critical gap in current analysis, motivating our in-depth study. 
%-------------------------------------------------------------------------------
\section{Benchmark Design}
%-------------------------------------------------------------------------------
\label{sec:benchmark}
\begin{figure}[t]
    \centering
    \includegraphics[width=1\linewidth]{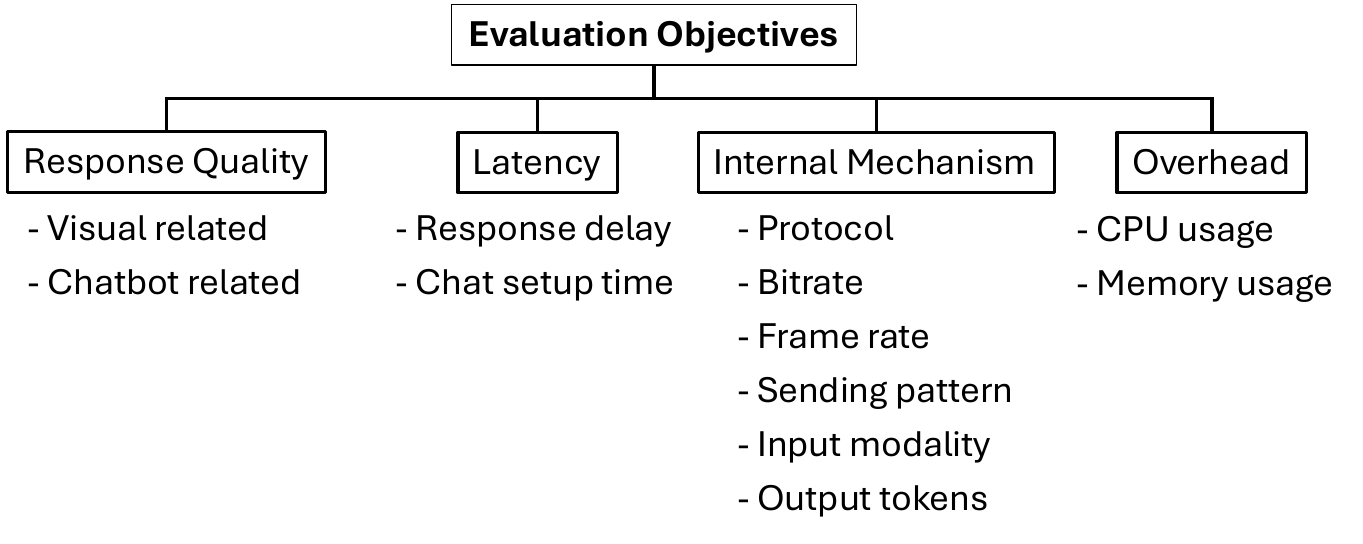}
    \caption{Evaluation objectives for AI video chat}
    \label{fig:metrics}
\end{figure}

To comprehensively evaluate the AI video chat application, it is crucial to assess both the user's experience and the system's internal performance. 
We want to know how good the user experience is -- the quality of AI and the latency -- using a series of objective metrics. 
This is just as in the traditional RTC applications where user experience contains the aspects of quality and latency. 
We also want to open up the box of the commercial AI video chat application to understand how it works and how many resources it consumes. 
Accordingly, we have designed and structured our specialized metrics into four key areas: quality, latency, internal mechanism, and system overhead (see Figure~\ref{fig:metrics} for details).

It is worth noting that in this section, we only present what the benchmark will measure.
We present the details of how we measure them (e.g., across different countries, at daytime and nighttime) later in \S\ref{sec:results}.
We will explain by different areas as below. 

% \subsection{Metric overview}
% The metrics are detailed in Figure~\ref{fig:metrics}. 
% For quality, we measure response quality (how good the AI response is) and perceptual quality (the speaking rate and AI response duration). 
% For latency, we measure response delay and video chat setup time. 
% For internal mechanism, we measure 

\subsection{Quality}
Quality is always the most important metric in AI video chatbot.
If the chatbot keeps answering nonsense or with a very limited knowledge (like Apple Siri in 2011~\cite{siri2011}), it will not be that widely used either.
We consequently evaluate the quality of AI video chatbot from the following perspectives:
\begin{itemize}
    \item Visual-related response quality. The chatbot should be able to handle typical visual tasks such as recognizing the objects.
    \item Chatbot-related response quality. AI video chatbot introduces new use cases such as memory recall, which also needs benchmarking.
    % \item Perceptual quality in response audio. Finally, the answer from the AI video chatbot should also be perceptually seamless to human beings.
\end{itemize}

We test the response quality by constructing a dataset that contains a series of videos and questions that cover all the aspects above.
We feed the videos and questions to the AI video chatbot and collect the response (testbed details in \S\ref{sec:testbed-design}).
We later determine how good the answer is with the ground truth. 
Below we present the details of these aspects.

\parahead{Visual-related response quality.}
Unlike text- or audio-based chat systems, the AI agent can not only perceive what the user sees but also operate in a RTC setting, delivering prompt responses akin to human video chat. 
Instead of receiving all video frames at once, the agents process a continuous video stream, consistent with real-time interaction. 
The dataset construction adheres to this streaming paradigm. 
We borrow the questions and answers from a visual task benchmark from the computer vision community, namely StreamingBench~\cite{lin2024streamingbench}. 
Our benchmark can also be extended to other datasets.

% We explain this in three aspects: 

% \noindent \textit{Why we chose this dataset?}
% There are two main reasons: 
% 1) Alignment with real-time requirements. 
% Unlike offline video understanding datasets that provide the entire video at once, it's specifically designed for real-time evaluation. 
% Questions are tied to timestamps, forcing the AI to respond using only the video content seen up to that moment, without access to future frames. 
% 2) Authoritativeness. It is the first large-scale, trustworthy benchmark created to evaluate the streaming video understanding ability of AI agents, making it an industry standard. 

% \noindent \textit {Why we need sampling?}
Nevertheless, it is non-trivial to test with all questions in the benchmark given that we're testing the AI video chatbot as a user and treat it as a black box.
Testing how an application works is different from testing LLMs using APIs or self-hosted models. 
% With the latter, efficiency can be achieved by directly feeding the required video frames and the corresponding question to the model. 
AI video chat applications must proceed in sync with the video: each test requires waiting for the entire video duration. 
Considering practical time limits, it is necessary to select a representative subset of the original dataset. 
Therefore, we have to only select a subset of questions and videos from the StreamingBench to adopt in our benchmark.

\textit {How to select the subset?}
Yet, it is non-trivial to select the subset while maintaining the results to be general.
We randomly selected videos to construct the subset following two principles: 
(1) the subset covers all evaluation aspects -- real-time visual understanding (e.g., counting), contextual understanding, omni-source understanding, and proactive output. 
% For example, the real-time visual understanding dimension includes sub-tasks such as counting, object perception, and spatial reasoning.  
(2) the subset spans a wide range of domains, such as life records, competitions, film and television, and education. 

We also need to preprocess the benchmark because the original dataset is composed of multiple-choice questions in text format, whereas our case requires audio-based input and direct conversion. 
We convert the multiple-choice questions into open-ended questions, ensuring each has a single, unambiguous answer. 
These open-ended questions are then transformed into audio.  
For omni-source understanding, which requires the video’s audio to evaluate the synchronous integration of visual and auditory inputs, we combine the video audio with the question audio. 
The video’s audio is attenuated to 20\% of the question’s volume, simulating environmental background sound~\cite{bg-music}.
For other task types, the input consists of the silent video paired only with the question audio. 
The final subset is summarized in Table~\ref{tab:dataset}, with each test video averaging 3.2 questions and 2 minutes 26 seconds in length. 

\begin{table}[t]
    \centering
    \begin{tabular}{l|cc}
        \hline
        \textbf{Category} & \textbf{Videos} & \textbf{Questions} \\ 
        \hline
        \multicolumn{3}{l}{Visual-related response quality} \\
        \hline
        \quad Real-time visual understanding & 15 & 54 \\
        \quad Contextual understanding & 6 & 24 \\
        \quad Omni-source understanding & 8 & 24 \\
        \quad Proactive output & 4 & 4 \\ 
        \hline
        \multicolumn{3}{l}{Chatbot-related response quality} \\
        \hline
        \quad Visual content memory & 3 & 3 \\ 
        \quad Visual named entity recognition & 15 & 15 \\
        \quad Math problem solving & 10 & 10 \\
        \hline
        \textbf{Total} & \textbf{61} & \textbf{134} \\ \hline
    \end{tabular}
    \caption{Overview of the constructed dataset}
    \label{tab:dataset}
\end{table}

\parahead{Chatbot-related response quality.}
The agent should be able to handle typical scenarios encountered in AI video chat, reflecting practical applications and user needs. 
Thus, in addition to evaluating the general capabilities of visual AI agents, we also aim to assess their performance in AI video chat–specific scenarios. 
To this end, we reviewed online reports, blogs, and posts, and summarized a set of common use cases. 
These scenarios primarily include: (a) identifying brands, dishes, etc.~\cite{obj-id-1,obj-id-2,obj-id-3}, (b) introducing tourist attractions~\cite{tour-1}, (c) assisting students in solving problems~\cite{prblm-1,prblm-2,prblm-3}, and (d) recalling information from previous video segments~\cite{mem-1}. 
We constructed a dataset by focusing on these typical, representative cases. 
% Consequently, we excluded recommendation-related scenarios (d) and 
We reframed tourist-attraction introductions (b) into recognition tasks (a). 
The final dataset consists of three categories of testing videos: (a) recognition, (b) problem solving, and (c) memory recall, with each video paired with a corresponding open-ended question. 
The first two (avg. 13.72 s) require AI to infer from the current frames, while the last (up to 10 min) requires inference over a longer history. 
The videos are sourced from YouTube~\cite{youtube-official}, Pexels~\cite{pexels-official}, and other online datasets. 
Their detailed definitions are as follows: 

\begin{figure}[t]
    \centering
    \includegraphics[width=1\linewidth]{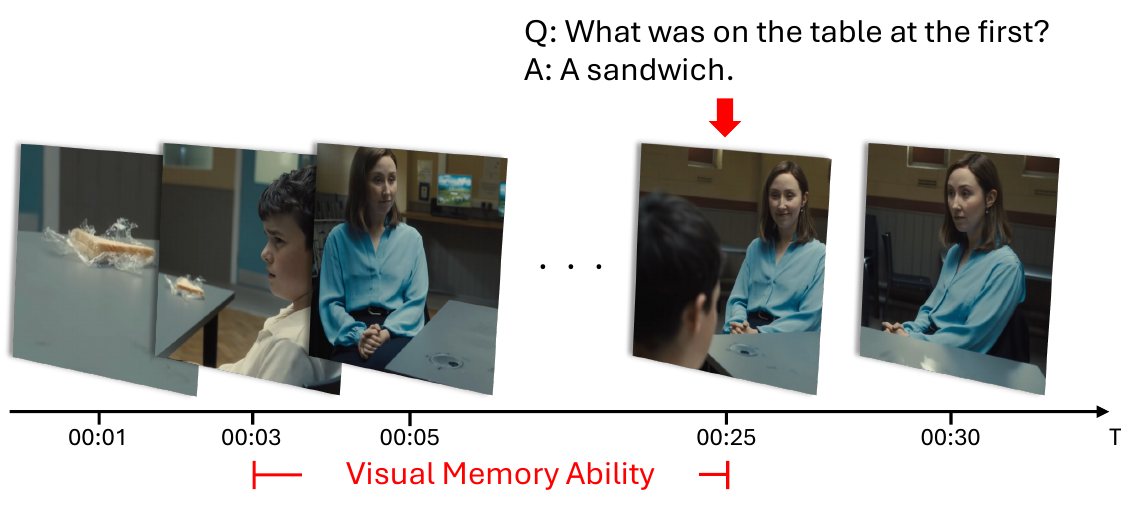}
    \caption{The definition of AI visual content memory}
    \label{fig:memory}
\end{figure}

\begin{itemize}
    \item Visual content memory. As shown in Figure~\ref{fig:memory}, this set of questions measure how long AI can recall visual details. 
We used 3 videos with distinct features at the beginning and asked about them at later timestamps (30 s, 1 min, 5 min). 
The intermediate questions before the final memory check are irrelevant to ensure that the text context does not reveal the visual information. 
\item Visual named entity recognition. 
We test the ability to identify fine-grained entities (brands, dishes, tourist attractions) rather than coarse objects (e.g., dogs vs. cats). 
We used 15 videos (5 per category), selecting the top-ranked items online~\cite{brands,foods,attractions}, and ensured diversity across categories.
\item Math problem solving.
We evaluate AI’s ability to solve math problems in video chat. 
We selected 10 videos (5 algebra, 5 geometry) with difficulty ranging from middle school to university.
\end{itemize}
We use \texttt{gTTS}~\cite{gTTS} Python module to convert text into speech for video chat input. 
Since video frames continue to stream during question playback, we ensure that the content remains consistent throughout each question’s audio so that the answer is not affected by its duration. 
For end-of-video questions, an additional blank screen segment is appended to ensure sufficient time for playback.

% \parahead{Perceptual quality.}
% Since AI audio is synthesized, we evaluate its perceptual quality using two straightforward metrics: speaking rate and response duration. 
% These capture how fast and how long the AI typically speaks, both of which directly affect user comprehension and experience. 
% Speaking rate (words per minute) reflects whether the AI’s pace is too fast to follow or too slow to sound natural. 
% Response duration (total seconds spoken per turn) indicates whether replies feel excessively long and overwhelming or too brief and unhelpful. 
% Together, these measures provide a direct assessment of the usability of synthesized speech.

% \parahead{Other Metrics.}
% We also define three additional metrics: video chat setup time, the delay from session initiation until the AI is ready; response duration, the length of the AI's audible reply; and speaking rate, the pace of its synthesized speech in words per minute (WPM). 

\subsection{Latency}
% Our evaluation of the AI video chat user experience is centered on two key dimensions -- latency and quality.
% User experience is primarily dictated by system responsiveness and the quality of its responses. 
% Excessive delays in chat setup or AI response times can lead to user abandonment, just as rapid but low-quality answers can diminish user satisfaction. 
% Thus, our approach synthesizes high-level evaluation metrics from related domains, including traditional video chat, AI text/audio agents, and video analytics. 
% These fields consistently focus on latency and quality.  
% Meanwhile, we also define evaluation metrics that are customized for AI video chat applications.
% Our latency metrics include video chat setup time and response delay. 
% Our quality metrics consist of response quality, speaking rate, and response duration. 
% We first highlight response delay and response quality, as they are two critical metrics for user experience in AI video chat. 
% A discussion of the other metrics follows. 

Latency significantly impacts user experience in the interaction. 
Even if the response quality is perfect, if it comes too late, users would rather find the answer themselves using text-based conversations with the chatbot. 
In AI video chat, it primarily consists of response delay and setup time, both of which can influence how smooth and natural the interaction feels. 
We explain these two aspects accordingly below.

\parahead{Response delay.}
We define response delay in the context of AI video chat to capture the latency directly perceived by the user. 
This metric differs from related concepts. 
In traditional video chat, delay typically refers to the end-to-end network latency of video transmission. 
In the LLM field, a similar metric is Time To First Token (TTFT), which measures the time until the first token of the response appears. 
In contrast, we measure response delay as the time interval from the moment the user stops speaking to the moment the AI begins its audible response. 
This specifically quantifies the conversational lag a user experiences while waiting for the AI to reply.

\parahead{Video chat setup time.}
We measure video chat setup time because a user's first impression is critical, and a long initial wait can cause them to abandon the session. 
This metric is defined as the total delay from when a user initiates the chat until the AI is fully connected and ready to interact. 
It reflects the time for establishing the connection and preparing the AI backend.

% \subsection{Internal mechanism}
% AI video chat involves a real-time video stream which is similar to traditional video chat. 
% However, it is also different since the replacement of human with AI. 
% Therefore, we measure the mechanism metrics that come from both traditional video chat (network protocols, data rate, and frame rate) and the LLM field (input modality). 

% \parahead{Input modality.}
% As a unique feature different from traditional video chat, AI video chat has LLM backend, and LLM will has its own characteristics. 
% Input modality is one significant feature, which means what kind of input the LLM actually process. 

% \parahead{Network protocol.}
% Since AI video chat involves a real-time video stream which is similar to traditional video chat, we want to investigate what protocols AI video chat adopts. 
% Thus, we identify the network protocols. 

% \parahead{Data rate.}
% Since AI video chat involves a uplink video \& audio stream and a downlink audio stream, we measure both side data rate. 

% \parahead{frame rate.}
% The user continuously transmit video stream to AI agent. 
% We measure the frame rate to see what level it is and whether it is different from traditional video chat. 

\subsection{Internal Mechanism}
AI video chat shares similarities with traditional video chat in transmitting real-time video streams but differs in replacing the human counterpart with an AI model.
Therefore, we would like to know how the shift of use cases affects the design choices of video streaming.
Accordingly, we measure a series of dimensions on the design choices of video streaming (network protocol, video bitrate, frame rate, and video packet sending pattern).
We further measure LLM-specific metrics (input modality and output tokens) to understand both how the LLM consumes the inputs and how verbose its replies are.

\parahead{Network protocol.}
For traditional RTC applications such as video conferencing, the RTP protocol is most widely used, e.g., in Zoom and Google Meet. 
Therefore, we want to investigate whether the commercial AI video chat applications follow the same practice.

\parahead{Video traffic patterns.}
For traditional video chat, applications usually require the bitrate and frame rate (e.g., 750\,Kbps and 24\,fps in Zoom~\cite{imc2022measurementzoom}) to be higher than a certain perceptual threshold. 
For video analytics, frame rate is allowed to decrease to several fps because the neural networks process the stream. 
Since the receiver of AI video chatbot applications has changed from human beings or simple neural networks to LLMs, we would like to know how the design considerations behind the bitrate and frame rate are affected as well. 
Furthermore, the sending pattern (e.g., pacing or bursty) might also be affected by the decision of bitrate and frame rate.
% Since users continuously transmit video streams to the AI agent, we evaluate the frame rate to determine its level and how it compares with traditional video chat. 

\parahead{Input modality.}
Unlike traditional video chat, AI video chat relies on an LLM backend, whose performance depends on the modality of inputs it processes (e.g., text, audio, video, or multimodal combinations). 
For example, some applications might transcribe the audio to text and then feed it into the LLM, while others directly preserve and process the raw audio signals. 
It is necessary to understand the differences in input modality across applications as well. 

\parahead{Output tokens.}
Beyond what the LLM ingests, the verbosity of what it returns is itself a user-facing characteristic.
Each AI reply is realized as a sequence of tokens that the application either reads aloud or renders on screen, so a longer reply directly translates into a longer turn for the user.
Different vendors may also tune this verbosity over time via prompt engineering or backend model swaps.
We therefore measure the average number of output tokens per AI reply for each application to characterize this LLM-side trait and to track how it shifts across measurement rounds.

\subsection{System Overhead}
Finally, system overhead on the client is also a critical but largely ignored aspect in the measurement, even if the LLM runs on the cloud.
Recall the usage experience of video conferencing applications such as Zoom -- users keep complaining about the overheating of their laptops or mobile phones~\cite{zoom-overheat}. 
Therefore, it is critical to measure the system overhead of video capturing and processing as well, which is an optimization direction for future work. 
To evaluate system overhead, we monitored the CPU and memory usage on the local device to measure the performance impact of each application. 

% %-------------------------------------------------------------------------------
% \section{Test Dataset Construction}
% %-------------------------------------------------------------------------------
% \label{sec:dataset}

%-------------------------------------------------------------------------------
\section{Testbed Design}
%-------------------------------------------------------------------------------
\label{sec:testbed-design}
We want our measurement to meet the two goals: 
\begin{itemize}
    \item Automated. Tests should run at scale with reproducible results. 
    \item Region-supported. The testbed should be deployable across multiple regions. 
\end{itemize}
% 1) Automated. Tests should run at scale with reproducible results. 
% 2) Region-supported. The testbed should be deployable across multiple regions. 
% 3) Application-agnostic data collection: This means that our measurement fits any AI video chat application. 

Based on these goals, a cloud-based testbed leveraging virtual devices is an ideal solution. 
% However, a limitation arises because \chatgpt and \grok providers do not permit the use of their applications on virtual devices~\cite{chatgpt-virtual}. 
However, a limitation arises because \chatgpt cannot run on virtual devices due to provider restrictions~\cite{chatgpt-virtual}. 
To address this, we set up both cloud and local testbeds, enabling measurements on applications wherever support is available. 
% Details of the two testbeds are presented in \S\ref{sec:testbeds} as follows. 

\subsection{Design Approach}
\begin{figure}[t]
    \centering
    \includegraphics[width=1\linewidth]{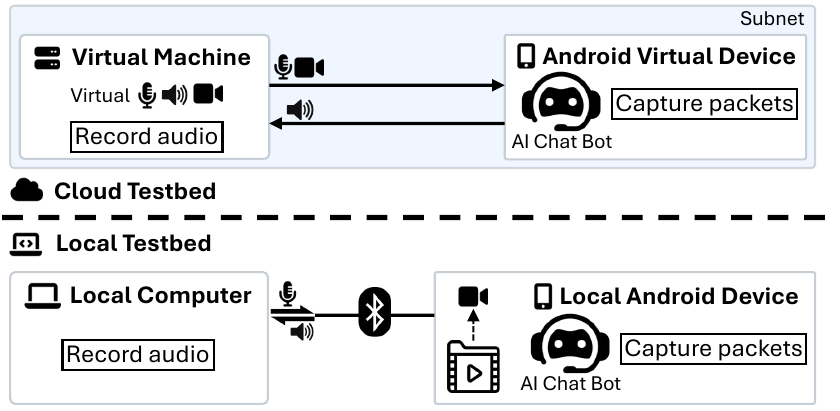}
    \caption{Overview of cloud and local testbeds}
    \label{fig:testbeds}
\end{figure}

% \label{sec:testbeds}
For both cloud and local testbeds, the key challenge lies in simulating audio and video streams to enable automated testing of video chat systems. 
We describe our approach for both testbeds as follows. 
The overview is shown in Figure~\ref{fig:testbeds}.

\noindent\textbullet \ \textbf{Cloud testbed.} 
The primary challenge in the cloud testbed is the absence of physical sensory devices on cloud-based virtual machines. 
To address this, we configure a virtual microphone and camera using a two-VM setup. 
The first VM runs an Android image on the Genymotion "Platform as a Service" (PaaS)~\cite{genymotion} to serve as an emulator. 
The second, a Linux VM, acts as a media source provider. 
Genymotion enables the Android emulator to treat the source provider's virtual microphone, speaker, and camera as its own by continuously streaming audio and video between the two. 
The media streams are managed as follows: 
\begin{itemize}
    \item Audio stream. We use \texttt{PulseAudio} to create a virtual microphone and speaker on the Linux VM. Audio files are played via \texttt{paplay}.
    % , while the virtual speaker captures the AI’s audio output. 
    \item Video stream. With \texttt{v4l2loopback} and \texttt{OBS}, we set up a virtual camera. Video files are streamed into it using \texttt{obsws\_python}. 
\end{itemize}
To ensure millisecond-level transmission latency, both VMs are located in the same subnet. 

\noindent\textbullet \ \textbf{Local testbed.} 
On the other hand, testing on physical devices presents its own challenges, primarily the need for a quiet environment. 
Such conditions are essential to prevent background noise from disrupting tests and to ensure accurate audio recording for latency measurements (see \S\ref{sec:setup}). 
To address this, we designed a local testbed that isolates the audio stream internally. 
Our setup connects a rooted Android device to a Linux computer via Bluetooth, redirecting the phone's microphone and speaker to the computer. 
We simulate the media streams as follows:
\begin{itemize}
    \item Audio stream. On the Linux machine, we create a virtual speaker and microphone and set them as the default devices. Audio files are then streamed using the same approach as in the cloud testbed. 
    \item Video stream. On the rooted Android device, video input is simulated using the \texttt{vcam} module, which creates a virtual camera fed by a local video file. 
\end{itemize}

Note that the \texttt{vcam} approach for video simulation does not currently function with Google \gemini. 
Table~\ref{tab:testbeds} summarizes the available testbeds for each application. 
We make sure each of these applications can run on at least one testbed.

We identify additional work required to ensure all testbeds run successfully.

\parahead{UI navigation.}
The whole process must be automated for every application. 
Since each application provides a different interface for starting a video call, enabling the camera, and ending the session, we use \texttt{adb} to simulate touch input events. 
This approach works for both cloud and local testbeds, as all clients run on Android devices. 

\begin{figure}[t]
    \centering
    \includegraphics[width=0.8\linewidth]{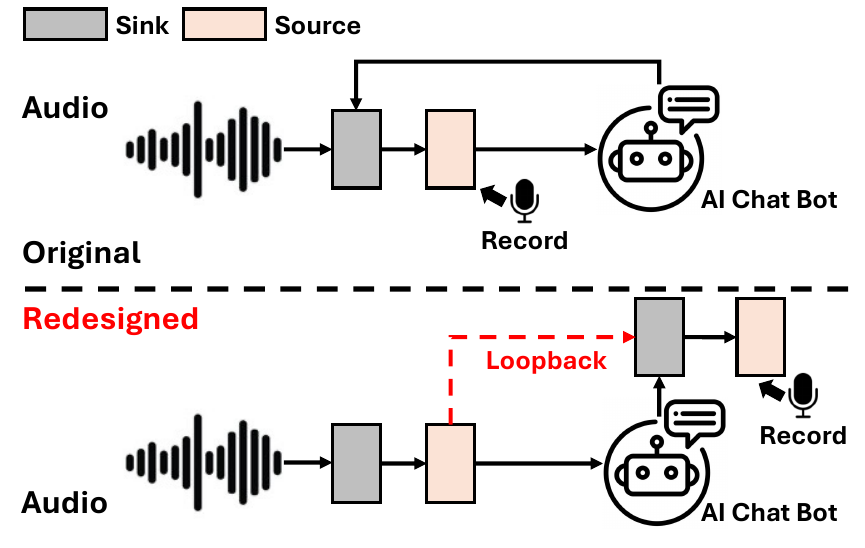}
    \caption{Reroute audio for echo cancellation}
    \label{fig:audio-route}
\end{figure}

\parahead{Echo cancellation.}
Both cloud and local testbeds face acoustic echo issues, where the AI hears its own responses and becomes interrupted. 
To address this, we redesign the audio routing using \texttt{pactl} commands. 
Figure~\ref{fig:audio-route} illustrates both the original and the redesigned audio paths. 
The terms source and sink refer to the audio input and output, respectively. 


\begin{table}[t]
  \centering
  \resizebox{\columnwidth}{!}{
  \begin{tabular}{l|cccccc}
      \hline
      & \chatgpt & \gemini & \grok & \doubao & \yuanbao & \qwen \\ 
      \hline
      Cloud &
      \ding{55} & $\checkmark$ & $\checkmark$ & $\checkmark$ & $\checkmark$ & $\checkmark$ \\ 
      % \hline
      Local &
      $\checkmark$ & \ding{55} & $\checkmark$ & $\checkmark$ & $\checkmark$ & $\checkmark$ \\ 
      \hline
  \end{tabular}
  }
  \caption{Available testbeds for each application}
  \label{tab:testbeds}
\end{table}

\begin{table}[t]
  \centering
  \begin{tabular}{l|ll}
      \hline
      & Testbed & Region(s) \\ 
      \hline
      \chatgpt & Local & Portugal \\ 
      % \hline
      \gemini & Cloud & US-East, US-West, Ireland, Hong Kong \\ 
      % \hline
      \grok & Cloud & US-East, US-West, Ireland, Hong Kong \\ 
      % \hline
      \doubao & Cloud & Hong Kong \\ 
      % \hline
      \yuanbao & Cloud & Hong Kong \\ 
      % \hline
      \qwen & Cloud & Hong Kong \\ 
      \hline
  \end{tabular}
  \caption{Each application's testbed and region(s)}
  \label{tab:setup}
\end{table}

% \subsection{Application-agnostic Data Collection}
% \label{sec:data-collect}
% Due to the black-box nature of AI video chat applications, we need to collect data in an application-agnostic way. 
% That is to say, testing should fit any AI video chat application and do not change the native behavior of these applications. 
% The data we can be categorized into two dimensions. 

% \parahead{User experience related.}
% We collect the metrics that will directly affect user experience. 
% AI response delay refers to the time interval from when the user finishes speaking to when the AI starts responding. 
% We also measure the quality of the AI response, namely whether it correctly answers the user's question.

% \parahead{Mechanism related.} 
% We measure both uplink (from user to AI) and downlink (from AI to user) data rates.
% We analyze frame rate and frame size based on the traffic packet analysis. We also analyze protocols. 

\begin{table*}[t]
    \centering
    \begin{threeparttable}
    \begin{tabular}{l|ccccccc}
        \hline
        % \toprule
        & \chatgpt & \gemini & \grok & \doubao & \yuanbao & \qwen & Human \\
        \hline
        \multicolumn{3}{l}{\textbf{Visual-related response quality}} \\
        \hline
        % \midrule
        \quad Real-time visual understanding & 47.69 & 69.15 & 29.96 & 60.84 & 27.48 & 46.05 & 91.46 \\
        \quad Contextual understanding & 35.87 & 51.57 & 28.74 & 53.70 & 29.75 & 34.42 & 91.40 \\
        \quad Omni-source understanding & 36.46 & 41.80 & 14.39 & 33.85 & 11.28 & 18.87 & 90.26 \\
        \quad Proactive output\textsuperscript{\dag} & 0\textbar{}0\textbar{}0 & 0\textbar{}0\textbar{}0 & 0\textbar{}0\textbar{}0 & 0\textbar{}25.00\textbar{}80.56 & 0\textbar{}0\textbar{}0 & 0\textbar{}0\textbar{}0 & 100\textbar{}100\textbar{}100 \\
        \hline
        \multicolumn{3}{l}{\textbf{Chatbot-related response quality}} \\
        \hline
        % \midrule
        \quad Visual content memory & \textgreater 10\,min & 7\textasciitilde8\,min & 0\textsuperscript{\ddag} & 1\textasciitilde2\,min & 0 & 0 & -- \\
        \quad Visual named entity recognition & 100 & 95.48 & 84.97 & 91.67 & 89.58 & 89.40 & -- \\
        \quad Math problem solving & 8.33 & 47.21 & 40.42 & 24.44 & 1.11 & 35.83 & -- \\
        \hline
        \textbf{Overall response quality\textsuperscript{$\star$}} & 46.49 & 62.04 & 34.11 & 55.17 & 30.11 & 43.03 & -- \\
        \hline
    \end{tabular}

    \begin{tablenotes}
        \item[\dag] within-1s \textbar{} within-5s \textbar{} within-10s. For each window, the score is the trigger rate, expressed as a percentage.
        \item[\ddag] \grok retains memory of prior visual content after the screen turns completely black.
        \item[$\star$] Overall score excludes \textit{proactive output} and \textit{visual content memory}.
    \end{tablenotes}
    \caption{Scores on the constructed datasets}
    \label{tab:scores}
    \end{threeparttable}
\end{table*}

\begin{table}[t]
    \centering
    \resizebox{\columnwidth}{!}{
    \begin{tabular}{l|cccccc}
        \hline
        & \chatgpt & \gemini & \grok & \doubao & \yuanbao & \qwen \\
        \hline
        Algebra & 16.67 & 66.94 & 54.63 & 42.22 & 2.22 & 49.72 \\
        Geometry & 0.00 & 27.48 & 26.22 & 6.67 & 0.00 & 21.94 \\
        \hline
    \end{tabular}
    }
    \caption{Scores for math problem solving (full mark: 100)}
    \label{tab:math-results}
\end{table}

\subsection{Experiment Setup}
\label{sec:setup}

% For the cloud testbed, due to regional restrictions, we use AWS Cloud~\cite{aws_cloud} to test \gemini outside China, and Alibaba Cloud~\cite{alibaba_cloud} to test \doubao and \yuanbao within China. 
For the cloud testbed, we use AWS Cloud~\cite{aws_cloud} to test the AI video chat applications. 
The video delay between two VMs (source provider: 4 vCPUs; Android emulator: Android 14 with 16 vCPUs) is approximately 300 ms. 
To measure this, the source provider plays a stopwatch video. 
We then capture the timestamps shown on (1) the source provider ($t_1$) and (2) the remote Android desktop as seen on the source provider ($t_2$). 
The video delay is computed as the difference between these two values, i.e., $t_1 - t_2$. 

For the local testbed, we perform experiments on a rooted Redmi Note 10 Pro (Android 13). 
The device connects to the Internet via WiFi with a bandwidth of about 100 Mbps. 
The audio transmission delay between the Android device and the computer over Bluetooth is about 200 ms. 
To measure it, we play audio on the Android phone while simultaneously recording on the Linux computer. 
The initial silence in the recording represents the transmission delay from Android to the computer ($t_1$). 
Similarly, we measure the delay from the computer back to the Android device ($t_2$). 
Thus, the total audio delay is given by $t_1 + t_2$.

% Below we explain how we measure the metrics in \S\ref{sec:benchmark}.

\parahead{Score computation.}
We transcribe each chat session with \texttt{Whisper} \cite{arxiv2022whisper} so that downstream grading runs on text rather than raw audio.
For all dimensions except \textit{visual content memory} and \textit{proactive output}, an LLM judge (Opus\,4.7~\cite{zheng2023judging}) compares each transcribed AI response against the ground truth on a $\{0, 1, 2\}$ scale (no match, partial, full match).
For each (app, question), we first pool every grade across all runs and all regions of that app and take the mean to obtain a per-question score in $[0, 2]$.
For each (app, dimension), we then average over the questions in that dimension and rescale by $\times 50$ to obtain a $[0, 100]$ score; the overall score per app is the mean over all (non-proactive) per-question scores, also rescaled by $\times 50$.
We exclude two dimensions from the overall score since they are not on the same scale.
\textit{Visual content memory} is reported as the longest elapsed time at which the agent can still recall an earlier visual fact (in seconds or minutes), so it is not on the same scale as the other dimensions.
\textit{Proactive output} is reported as a triple $x\,|\,y\,|\,z$, where $x$, $y$, $z$ are the percentages of trials in which the agent fired the requested output within $1$\,s, $5$\,s, and $10$\,s of the trigger event, respectively.
The triggers are visual events that occur naturally in the source video (e.g., the moment a basketball player takes their first shot).

\parahead{Latency measurement.}
In both cloud and local testbeds, audio packets arrive at the client before the AI begins speaking, so packet timing alone cannot mark the start of the AI's voice.
We instead record the entire chat session with \texttt{parecord} and analyze the audio with \texttt{librosa} to detect the user-finish and AI-start onsets, and report their interval as the response delay.
For video chat setup time, we screen-record the Android device and measure the duration from app launch to the first AI utterance.

\parahead{Internal mechanism.}
We capture network traffic with \texttt{tcpdump} on the Android device and analyze the traces offline.
Server locations are resolved with \texttt{MaxMind}~\cite{maxmind} and \texttt{ipinfo.io}~\cite{ipinfo_io}.
We probe each agent's input modality with a series of targeted prompts.
For output verbosity, we count tokens in the Whisper transcripts with the \texttt{cl100k\_base} tokenizer, which handles English and Chinese consistently (roughly 3--4 chars per token for English and one token per Chinese character).

\parahead{System overhead.}
We track CPU and memory usage on the Android devices with the \texttt{top} command throughout each session.

Table~\ref{tab:setup} lists, for each application, the testbed and the deployment regions used in our primary experiments.
For \grok, \doubao, \yuanbao, and \qwen, we additionally repeated the experiments on the local Redmi testbed as a sanity check against cloud-environment artifacts.

We ran the experiments in two rounds, the first in August 2025 and the second in March 2026, which lets us track how each application evolved over the intervening months.
The 2025-08 round covered \chatgpt, \gemini, \grok, \doubao, and \yuanbao.
The 2026-03 round covered \gemini, \grok, \doubao, \yuanbao, and \qwen: \chatgpt was excluded because OpenAI currently doesn't support its video chat feature in this round~\cite{chatgpt_video_removed}, and \qwen was added because its video chat product launched only after the first round.
Within a round, every (application, region) pair was tested for at least two days.
A single pass through our dataset takes about six hours per region.

All test videos were rendered at $1280\times720$, the maximum camera resolution supported by the Genymotion cloud emulator we use.
At the start of each video, we open a new chat and clear any prior conversation, so that the AI cannot rely on memory carried over from a previous test.

%-------------------------------------------------------------------------------
\section{Measurement Results}
%-------------------------------------------------------------------------------
\label{sec:results}

In this section, we first present an overview of performance in~\S\ref{sec:perf_overview}, followed by quality-related performance in~\S\ref{sec:quality-result}, latency-related performance in~\S\ref{sec:latency-result}, internal mechanisms in~\S\ref{sec:mechanism-result}, and system overhead in~\S\ref{sec:overhead-result}.
To understand the behavior of AI video chat applications in highly bitrate-constrained scenarios, we also conducted controlled testbed experiments, which are detailed in~\S\ref{sec:low-rate-result}.
Unless noted otherwise, every number reported in this section is pooled across all regions of an application and taken from its most recent measurement round -- 2026-03 for \gemini, \grok, \doubao, \yuanbao, and \qwen, and 2025-08 for \chatgpt, whose video chat feature was discontinued before the 2026-03 round~\cite{chatgpt_video_removed}; error bars in figures (where present) represent the standard deviation across runs.

\subsection{Performance Overview}
\label{sec:perf_overview}

\begin{figure}[t]
    \centering
    \includegraphics[width=0.75\linewidth]{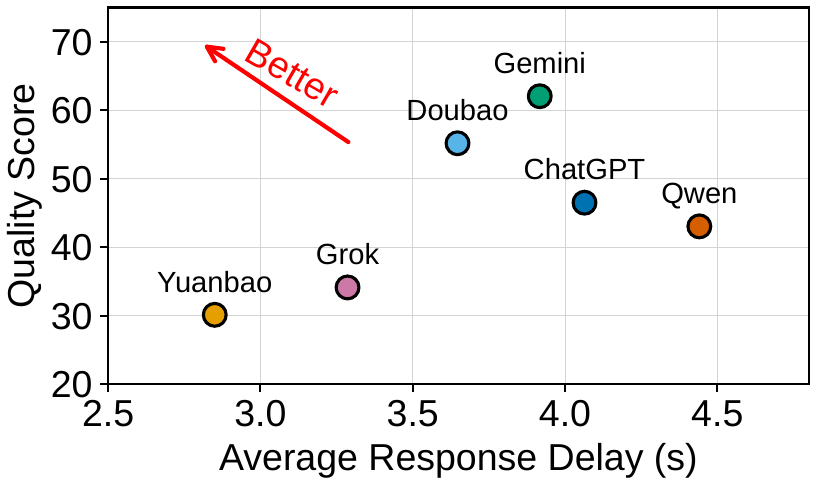}
    \caption{Overall performance of AI video chat applications}
    \label{fig:tradeoff}
\end{figure}

% \begin{figure}[t]
%     \centering
%     \includegraphics[width=0.75\linewidth]{figures/score-change.pdf}
%     \caption{Video chat agent capability over time}
%     \label{fig:score_change}
% \end{figure}

\begin{table}[t]
    \centering
    \begin{threeparttable}
    \resizebox{\columnwidth}{!}{
    \begin{tabular}{c|cccccc}
        \hline
        & \chatgpt & \gemini & \grok & \doubao & \yuanbao & \qwen \\
        \hline
        \multicolumn{3}{l}{\textbf{Quality score}} \\
        \hline
        \quad 2025-08 & 46.49 & 44.32  & 36.42  & 54.22  & 21.99 & --    \\
        \quad 2026-03 & -- & 62.04  & 34.11  & 55.17  & 30.11 & 43.03 \\
        \quad $\Delta$ & --    & +17.72 & --2.31 & +0.95  & +8.12 & --    \\
        \hline
        \multicolumn{3}{l}{\textbf{Response delay (s)}} \\
        \hline
        \quad 2025-08 & 4.06  & 3.30  & 2.71  & 3.67   & 2.42  & --   \\
        \quad 2026-03 & --    & 3.92  & 3.29  & 3.65   & 2.85  & 4.44 \\
        \quad $\Delta$ & --    & +0.62 & +0.58 & --0.02 & +0.43 & --   \\
        \hline
    \end{tabular}
    }
    \caption{Average quality score and response delay across two rounds.}
    \label{tab:quality-change}
    \end{threeparttable}
\end{table}

Figure~\ref{fig:tradeoff} illustrates the overall performance of the six AI video chat applications along two key dimensions: response delay and response quality score.
The quality score excludes \textit{visual content memory} (reported in seconds) and \textit{proactive output} (reported as a three-window triple), since they are not on the same scale as the other dimensions.

Our evaluation reveals a clear trade-off between response quality and delay across the six applications, with no single agent leading on both axes.
\yuanbao and \grok occupy the lower-left of the figure, delivering the fastest responses (2.85\,s and 3.29\,s) but at the cost of the two lowest quality scores (30.11 and 34.11).
\gemini, \doubao, and \chatgpt occupy the upper-right, producing the three highest quality scores (62.04, 55.17, 46.49) at noticeably longer delays (3.92\,s, 3.65\,s, 4.06\,s).
\qwen sits off the trade-off frontier rather than on it: it has the longest delay of any agent (4.44\,s) but only a mid-tier quality score (43.03), so it pays the latency cost of a high-quality system without delivering the corresponding quality.
The fact that the higher-quality agents are also the slower ones is consistent with more sophisticated processing workflows (e.g., deeper reasoning, more reasoning steps) being a prerequisite for higher response quality, but \qwen's position shows that high latency alone does not guarantee high quality.

\parahead{Quality and delay change between rounds.}
Table~\ref{tab:quality-change} reports per-app means for the two measurement rounds.
\chatgpt was only present in 2025-08 (its video chat feature was not supported in the 2026-03 round~\cite{chatgpt_video_removed}) and \qwen was only present in 2026-03 (its video chat product launched between rounds), so the round-over-round comparison is restricted to \gemini, \grok, \doubao, and \yuanbao.
Three of the four agents improved in quality (\gemini\,+17.72, \yuanbao\,+8.12, \doubao\,+0.95) and one declined slightly (\grok\,--2.31); response delay grew for the same three (\gemini\,+0.62\,s, \grok\,+0.58\,s, \yuanbao\,+0.43\,s) and was effectively flat for \doubao (--0.02\,s).
Two patterns are worth noting.
First, the agents that gained the most quality (\gemini, \yuanbao) also became slower, suggesting that recent vendor updates buy reasoning depth at the cost of latency rather than improving both axes at once -- AI quality progress in this product category is not free.
Second, \grok stands out as the only agent whose quality dropped while its delay grew, and this coincides with a roughly $70\%$ jump in average output tokens per reply (\S\ref{sec:mechanism-result}); longer replies appear to dilute rather than reinforce its response quality.
\doubao's near-stationary trajectory on both axes hints that it has converged on a stable operating point in this product cycle.

% This finding highlights a challenge for future optimization: developers must carefully balance the competing demands of response speed and quality. 

\subsection{AI Quality Analysis}
\label{sec:quality-result}
We evaluated quality-related performance, focusing on visual-related (\S\ref{sec:visual-related}) and chatbot-related (\S\ref{sec:chatbot-related}).
We measured the applications on our constructed datasets. 
The results are shown in Table~\ref{tab:scores}. 
In this section, we analyze what's behind those scores. 

\subsubsection{Visual-related response quality.}
\label{sec:visual-related}
% In terms of \textit{real-time visual understanding} and \textit{contextual understanding}, \doubao achieved the highest score of 65.28 and 45.97 -- over 3 times that of \yuanbao, which scored the lowest at 19.44 and 16.67. 
% However, this top AI score remains far below the human performance reported in StreamingBench~\cite{nips2024streambench}, which exceeds 90.  
% This gap highlights substantial room for improvement in AI’s general ability to understand streaming video.
AI's general ability to understand streaming video still has substantial room for improvement.
On every visual sub-dimension, the highest AI score lies well below the human reference on StreamingBench~\cite{lin2024streamingbench}, where humans consistently exceed 90.
On \textit{real-time visual understanding}, the strongest agent (\gemini, 69.15) trails the human reference (91.46) by more than 22 points; on \textit{contextual understanding} the best agent is \doubao at 53.70 against a human 91.40; and on \textit{omni-source understanding} the field collapses further -- the best agent (\gemini, 41.80) is more than 48 points below human, and three agents (\grok, \yuanbao, \qwen) score below 20.
Each agent also degrades from its own real-time score as the task moves to contextual and then to omni-source understanding -- \gemini, for example, loses about 18 points to contextual and 27 points to omni-source.
This pattern indicates that the central difficulty is not recognizing isolated visual content but reasoning over visual context and integrating audio with vision under real-time constraints.

\textit{Proactive output} remains the weakest dimension across the field, but is no longer uniformly absent.
A typical instruction is of the form: \textit{``When the player takes their first shot, say `First Shot Taken.'\,''}.
In our previous measurement round, every agent confirmed the instruction (\textit{``Yes, I will.''}) and then never fired at the trigger event -- all six agents scored $0\,|\,0\,|\,0$.
In the latest round, \doubao is the only agent showing emerging capability: it triggers within 5\,s on 25.00\% of trials and within 10\,s on 80.56\%, but still on 0\% of trials within 1\,s.
The other five agents continue to behave passively, responding only after detecting user speech rather than initiating outputs when the trigger event appears in the video.
This pattern suggests that proactive temporal alignment with the visual stream -- not just visual recognition itself -- is the next frontier for video chat agents.

\subsubsection{Chatbot-related response quality.}
\label{sec:chatbot-related}
% To assess the contextual memory of each AI agent, we measured the duration over which they could recall previously presented visual information. 
% The results varied widely: \chatgpt exhibited the longest retention period at over ten minutes, while \yuanbao demonstrated a complete inability to recall information once it was no longer in the immediate video frame. 
We assessed \textit{visual content memory} by measuring how long each agent retained earlier visual information, and found very wide variation across the six agents (Table~\ref{tab:scores}).
\chatgpt showed the longest retention at 10+ minutes, followed by \gemini at 7--8 minutes and \doubao at 1--2 minutes.
At the other end, \yuanbao and \qwen show no recall at all once the relevant content leaves the current video frame, and \grok also reports zero in the standard setting (with the blank-screen exception discussed below).
We hypothesize that \yuanbao and \qwen employ a speech-gated mechanism in which video processing is triggered only by user speech.
However, despite only processing video intermittently, both clients continuously transmit the video stream over the network (\S\ref{sec:mechanism-result}).
This mismatch between data transmission and computation leads to unnecessary bandwidth consumption.
An ideal architecture should synchronize streaming with processing, pausing transmission during periods of user silence.

% In memory ablity part, we also found an interesing finding. 
% As shown in Table~\ref{tab:scores}, \grok has no memory for most cases we test. It do have memory when the screen becomes totally blank and then we asked whether it remember things before. 
% We infer that black screen has no new information so \grok will maintain its original memory and can remember things like 10 seconds ago. 
% However, in most cases, its memory will be overwrite with the current frames which contains large information. 
% \grok is different with \yuanbao since their memory are both almost 0. \grok do has memory but will be easily overwritten, but \yuanbao has no memory at all.  

% In the memory evaluation, we also found that \grok generally lacks memory, except when the screen is blank, in which case it can recall prior content. This suggests that new visual input quickly overwrites its memory. Compared with \yuanbao, which shows no memory at all, \grok has a fragile but overwrite-prone memory. 
We also observed an interesting phenomenon in the memory evaluation. 
As shown in Table~\ref{tab:scores}, \grok generally exhibited no memory across most test cases. 
However, when the screen turned completely blank and we asked questions about earlier content, the model was able to recall events from up to 10 seconds prior. 
We infer that the absence of new visual input during a blank screen allows \grok to retain its prior memory, whereas in typical scenarios, incoming frames with rich information quickly overwrite it. 
Compared with \yuanbao and \qwen, all models exhibit near-zero memory overall, but their mechanisms differ: \grok shows a fragile memory that is easily overwritten, whereas the other two has no memory capacity as it does not process these video frames at all.

By contrast, all six agents perform strongly on \textit{visual named entity recognition}, with every score exceeding 80 (\chatgpt reaches 100; the lowest, \grok, still reaches 84.97).
This indicates that, when a question can be resolved by mapping a recognizable on-screen entity to external knowledge, the agents reliably retrieve that knowledge.
The contrast with the much lower scores on \textit{contextual} and \textit{omni-source understanding} confirms that the bottleneck for these agents is not entity-level recognition but reasoning over visual context and integrating audio with vision.
% Notably, when contrasted with the substantially lower scores observed in the StreamingBench subset, a clear pattern emerges: tasks demanding reasoning over contextual information or differentiation among visually redundant inputs result in significantly degraded performance. 
% By comparison, visual named entity recognition poses no such challenges.

% \begin{table}[t]
%     \centering
%     \begin{tabular}{l|l l}
%         \hline
%         \  & Algebra & Geometry \\ \hline
%         \chatgpt & 50.00 & 0 \\ \hline
%         \gemini & 70.00 & 0 \\ \hline
%         \grok & 0 & 0 \\ \hline
%         \doubao & 80 & 80 \\ \hline
%         \yuanbao & 50 & 0 \\ \hline
%     \end{tabular}
%     \caption{Scores for math problem solving}
%     \label{tab:math-results}
% \end{table}

The results in \textit{math problem solving} reveal a broad limitation across all six agents.
Against a customary 60-point passing threshold, no agent reaches a passing score on this category in Table~\ref{tab:scores}: even the strongest, \gemini at 47.21, falls short, while \yuanbao and \chatgpt sit at the bottom (1.11 and 8.33).
Within this overall weakness, geometry is consistently the harder of the two skills (Table~\ref{tab:math-results}): every agent scores lower on geometry than on its own algebra, and three of the six (\chatgpt, \yuanbao, \doubao) score below 10 on geometry, with \chatgpt and \yuanbao failing to produce a single correct geometry solution.
We attribute this gap to the different cognitive skills required: geometry necessitates visual-spatial reasoning over shapes, whereas algebra primarily relies on symbolic processing, which is a comparatively simpler task for these models.
These findings suggest that today's AI video-chat agents remain limited in their ability to assist students with mathematical problem solving, particularly in geometry.

\begin{figure*}[t]
  \centering
    \begin{minipage}[c]{0.31\textwidth}
        \centering
        \includegraphics[width=\textwidth]{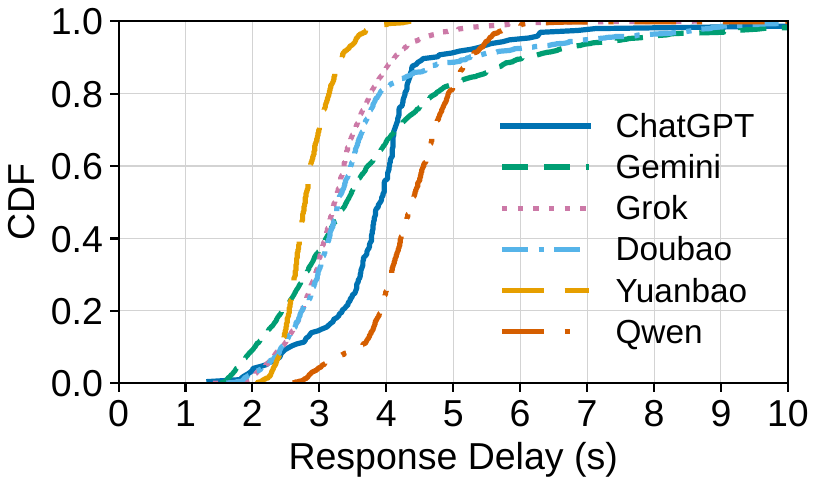}
        \caption{Response delay across apps}
        \label{fig:delay-app}
    \end{minipage}
    \hfill
    \begin{minipage}[c]{0.31\textwidth}
        \centering
        \includegraphics[width=\textwidth]{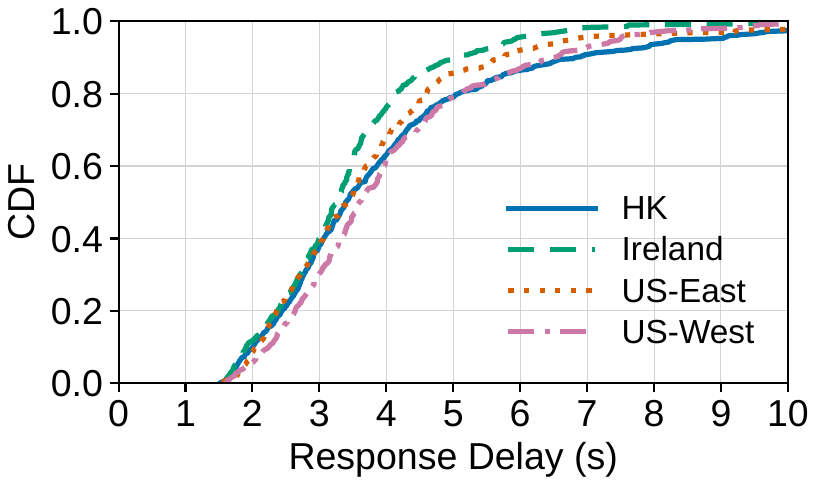}
        \caption{Delay across regions (\gemini)}
        \label{fig:delay-regions-gemini}
    \end{minipage}
    \hfill
    \begin{minipage}[c]{0.31\textwidth}
    \centering
    \includegraphics[width=\textwidth]{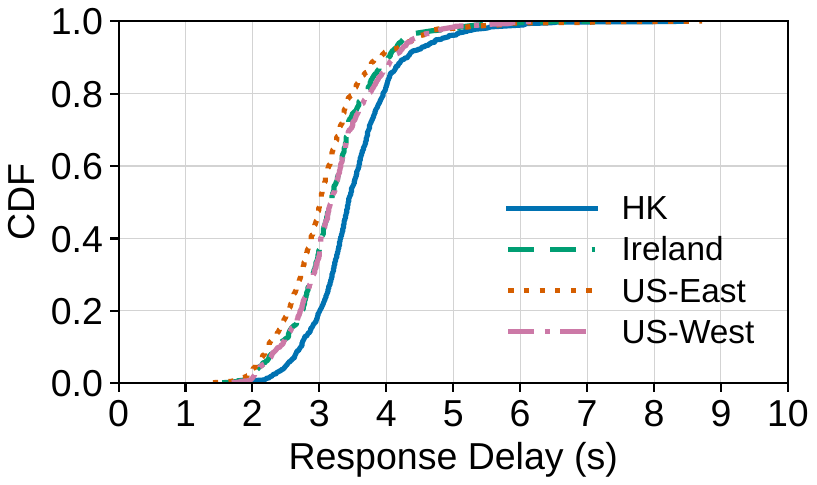}
		\caption{Delay across regions (\grok)}
		\label{fig:delay-regions-grok}
    \end{minipage}
\end{figure*}

\begin{figure*}[t]
  \centering
    \begin{minipage}[c]{0.31\textwidth}
        \centering
        \includegraphics[width=\textwidth]{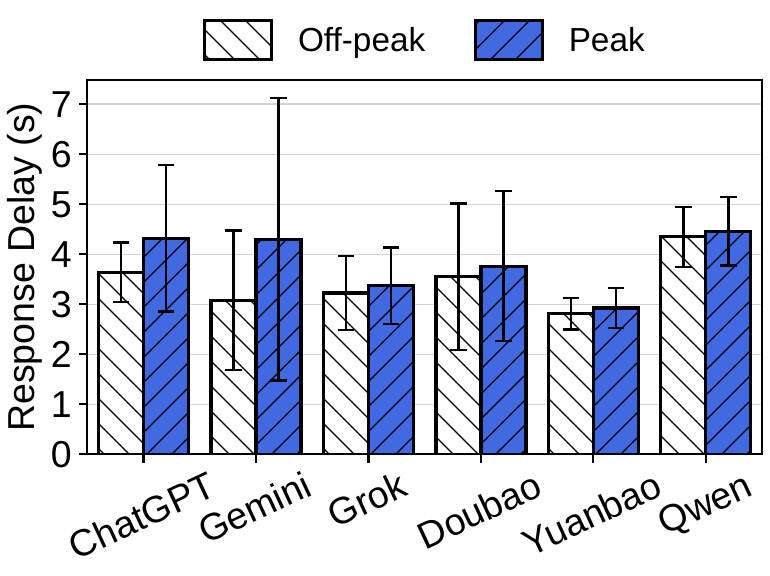}
        \caption{Comparison of peak and off-peak hours}
        \label{fig:peakhour}
    \end{minipage}
    \hfill
    \begin{minipage}[c]{0.31\textwidth}
        \centering
        \includegraphics[width=\textwidth]
        {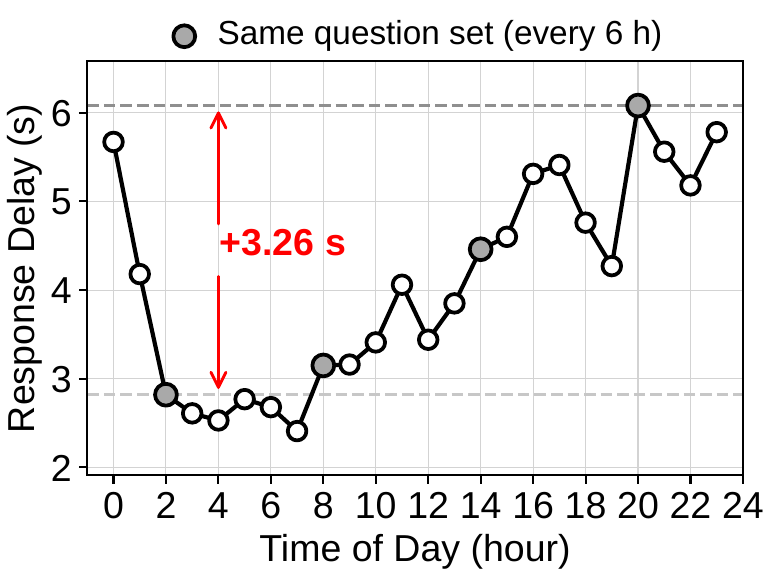}
        \caption{Response delay throughout the day (\gemini, Hong Kong)}
        \label{fig:delay-day}
    \end{minipage}
    \hfill
    \begin{minipage}[c]{0.31\textwidth}
    \centering
    \includegraphics[width=\textwidth]{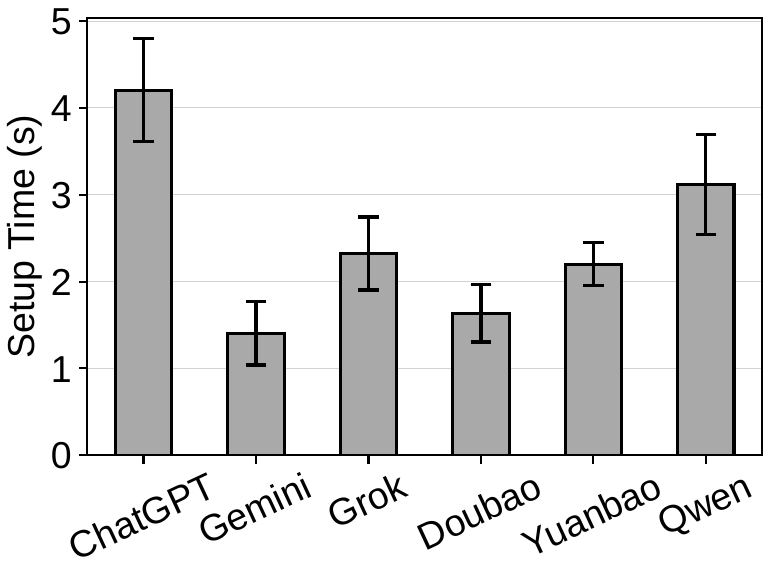}
		\caption{Video chat setup time}
		\label{fig:setup-time}
    \end{minipage}
\end{figure*}

\subsection{Latency Analysis}
\label{sec:latency-result}

In this section, we conduct an analysis and comparison of latency-related metrics, including response delay (\S\ref{sec:response-delay-result}) and video chat setup time (\S\ref{sec:setup-result}).
% First, we compare the response latency across different AI video chat applications, followed by an investigation into the various factors that influence AI response latency.

\subsubsection{Response Delay}
\label{sec:response-delay-result}

% \begin{figure}[t]
%     \centering
%     \includegraphics[width=0.7\linewidth]
%     {figures/delay-apps.pdf}
%     \caption{Response delay across apps}
%     \label{fig:delay-app}
% \end{figure}

We measured the response delay and find the following three points.

\parahead{High response delay across all applications.}
As shown in Figure~\ref{fig:delay-app}, the average AI response delay exceeds 2.8 seconds for every application.
Given that video chat requires a sub-second response interval~\cite{itu-t_g114_2003}, this delay level is insufficient to meet the demands of current AI-powered video chat scenarios.
Among these applications, \yuanbao achieves the lowest average AI response delay at 2.85\,s, while \qwen exhibits the longest at 4.44\,s -- a 1.56$\times$ gap between the two.
For the remaining applications, \chatgpt averages 4.06\,s, \gemini 3.92\,s, \doubao 3.65\,s, and \grok 3.29\,s.
The median response latency follows the same ranking.
Notably, \yuanbao, \grok, and \qwen are more stable than the others: their p95--mean gaps stay within 1.2\,s, whereas the heavier-tailed three (\gemini, \chatgpt, \doubao) all stretch to p99 above 9 seconds (12.5\,s, 11.5\,s, and 9.4\,s respectively).
\gemini in particular exhibits the widest p95 (7.59\,s), even though its mean is only moderate.

Then, what contributes to the increase in response delays? 
Beyond the inference time of the AI agent itself, our analysis points to two more potential causes, which we elaborate on below.

% \begin{figure*}[t]
%     \centering

%     % Left 1: apps
%     \begin{minipage}[b]{0.31\textwidth}
%         \centering
%         \includegraphics[width=\linewidth]{figures/delay-apps.pdf}
%         \caption{Response delay across apps}
%         \label{fig:delay-app}
%     \end{minipage}
%     \hfill
%     % Left 2 and Left 3: grouped as one figure
%     \begin{minipage}[b]{0.65\textwidth}
%         \centering

%         \begin{subfigure}[b]{0.48\linewidth}
%             \centering
%             \includegraphics[width=\linewidth]{figures/delay-regions-gemini.pdf}
%             \caption{\gemini}
%             \label{fig:delay-regions-gemini}
%         \end{subfigure}
%         \hfill
%         \begin{subfigure}[b]{0.48\linewidth}
%             \centering
%             \includegraphics[width=\linewidth]{figures/delay-regions-grok.pdf}
%             \caption{\grok}
%             \label{fig:delay-regions-grok}
%         \end{subfigure}

%         \caption{Response delay across different regions}
%         \label{fig:delay-regions}
%     \end{minipage}

% \end{figure*}

\parahead{Client locations can affect response delay.}
Leveraging cloud services, we measured the AI response delays of \gemini and \grok across different regions. 
Specifically, they were tested in four regions: US-East (Virginia), US-West (Oregon), Ireland, and Hong Kong.
The corresponding results are presented in Figure~\ref{fig:delay-regions-gemini} and \ref{fig:delay-regions-grok}, respectively.

\begin{itemize}
    \item \gemini. Response delay varies by region: Ireland is clearly the lowest at 3.46\,s, while HK (4.12\,s), US-West (4.07\,s), and US-East (3.90\,s) all cluster around 4.0\,s and are within \textasciitilde220\,ms of each other. The Ireland--vs.--rest gap is therefore the salient one (\textasciitilde500--660\,ms). Since clients accessed servers within their own region, client-server geographic distance can be ruled out; we therefore hypothesize that the Ireland \gemini service bears lighter load than the other three, which may explain its shorter delays.
    \item \grok. HK has the highest delay among the four regions (3.52\,s, vs.\ 3.08--3.24\,s for US-East, Ireland, and US-West). We believe this is because the \grok server is located in the US, so HK clients incur extra cross-Pacific RTT on top of the inference time.
    % \item \yuanbao. A \textasciitilde300 ms delay gap was found between China-North and China-South. Unlike \gemini, this discrepancy may stem from server allocation: our analysis showed Beijing clients were routed to servers in China-East (Shanghai) rather than to local Beijing servers. This cross-region assignment means Beijing-initiated requests must compete with Shanghai’s for server resources, possibly contributing to longer delays for Beijing clients.
    % \item \doubao. In contrast, no significant response delay difference was observed for \doubao between the two tested regions. Additionally, verification confirmed \doubao clients connected to local-region servers. The lack of a notable delay gap thus suggests \doubao operates under similar load conditions across these two regions.
\end{itemize}

\parahead{Request scheduling contributes to response delay.}
As previously noted, request burden exerts an impact on AI response delay.
To approximate the extent of this contribution, we measured the AI response delays of all applications during two time windows of the day in local time: peak hours (14:00--21:00) and off-peak hours (02:00--09:00), with identical questions posed in both periods.
As illustrated in Figure~\ref{fig:peakhour}, all six applications run slower during peak hours than during off-peak hours, but the magnitude of the gap differs widely.
\gemini exhibits the largest peak-hour penalty: its peak-hour mean (4.30\,s) is 40\% higher than its off-peak mean (3.08\,s).
\chatgpt is next at +19\% (4.32\,s vs 3.64\,s), while \doubao (+6\%), \grok (+4\%), \yuanbao (+4\%), and \qwen (+3\%) are nearly flat across the two windows.
We also provide a per-hour view of the largest-gap application.
Figure~\ref{fig:delay-day} illustrates how \gemini's HK response delay varies over a 24-hour period.
One full pass through our question set takes \textasciitilde6 hours, so the four sample points spaced 6\,h apart (02:00, 08:00, 14:00, 20:00) are queried with the same set of questions and are therefore directly comparable.
Within those comparable points, the peak hour (20:00, 6.08\,s) is 3.26\,s slower than the off-peak hour (02:00, 2.82\,s).
In other words, request scheduling accounted for at least 54\% (3.26/6.08) of the peak-hour response delay for this application. 
% This finding underscores the significant impact of scheduling on response latency. 
% While the inference time of AI models remains a primary contributor to overall latency, scheduling also accounts for a substantial portion of the delay -- highlighting its non-negligible role in the end-to-end latency performance.

\subsubsection{Video Chat Setup Time.}
\label{sec:setup-result}
We analyzed screen recordings of each application, measuring the time interval from the initial button click to the moment the service reached a "ready" state.
This "ready" state was defined by distinct visual cues, such as the activation of the camera view (evidenced by a transition from dark to bright) or the appearance of on-screen prompts like "Listening."
The results, presented in Figure~\ref{fig:setup-time}, show a roughly 3$\times$ spread across applications.
\chatgpt has the longest setup at \textasciitilde4.20\,s, followed by \qwen at \textasciitilde3.12\,s.
\grok and \yuanbao are in the middle (2.32\,s and 2.20\,s respectively), while \doubao (1.63\,s) and \gemini (1.41\,s) reach the "ready" state fastest -- under 1.7\,s.

% \begin{figure}[t]
%     \centering
%     \includegraphics[width=0.7\linewidth]{figures/schedule.pdf}
%     \caption{Response delay throughout the day (\doubao)}
%     \label{fig:delay-day}
% \end{figure}

\subsection{Internal Mechanism}
\label{sec:mechanism-result}

In this section, we investigate the internal mechanisms of each application.
We conduct a detailed analysis of captured network packets to evaluate key system performance metrics, including protocols, bitrate, frame rate, and video packet sending pattern.
We further examine two LLM-specific characteristics: the input modality of the underlying model -- probed via a series of targeted prompts -- and the verbosity of the AI's response, measured as the average number of output tokens per reply.
% -- consistent with the methodology discussed in \S~\ref{sec:background}
All findings from this section are summarized in Table~\ref{tab:mechanism}.

\parahead{Network protocols.}
In terms of application-layer protocols, we observe heterogeneity across different AI video chat systems.
Specifically, \chatgpt, \grok, \doubao, \yuanbao, and \qwen all rely on the RTP/RTCP stack -- a protocol suite widely adopted in RTC systems for media transport (RTP) and feedback / control (RTCP), typically multiplexed on the same UDP port (RTCP-mux, RFC~5761).
By contrast, \gemini employs QUIC at both the transport and application layers, which provides its own congestion control and reliability without a separate control protocol.

Although every RTP-based application carries audio and video on independent SSRCs (Synchronization Source Identifiers), the number of SSRCs per session varies (Table~\ref{tab:mechanism}).
\chatgpt, \yuanbao, and \qwen use the minimum of two SSRCs (one audio + one video), while \doubao occasionally adds short-lived auxiliary streams (2--4 SSRCs total).
\grok stands out with five concurrent uplink SSRCs: one audio, one always-on baseline simulcast layer at \textasciitilde1\,fps, two higher-resolution simulcast layers, and one RTX (retransmission) stream.
This is consistent with standard WebRTC simulcast (RFC~8853) and RTX (RFC~4588), but is unusual in our sample set -- the other RTP-based applications transmit only a single video layer.
Doubao's RTP packetization is also non-standard: each packet is prefixed with a fixed 2-byte header (\texttt{0xca 00}) before the RFC-3550 RTP header, requiring a custom parser to extract media metadata from the wire.

% \parahead{Number of connections.}

\begin{figure*}[t]
  \centering
  %---- 三个小图水平排开 ----%
  \begin{minipage}[c]{0.31\textwidth}
    \centering
    \includegraphics[width=\textwidth]{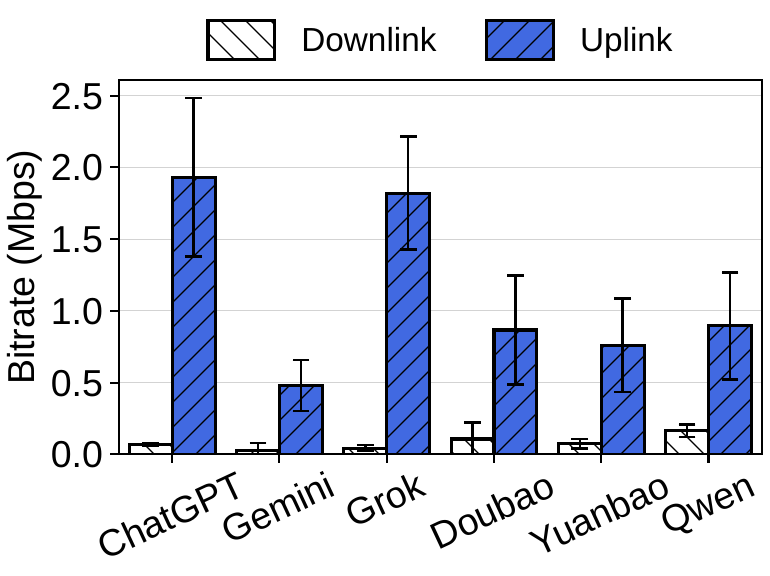}
    \caption{Video bitrates across apps}
    \label{fig:video-bitrate}
  \end{minipage}
  \hfill
  \begin{minipage}[c]{0.31\textwidth}
    \centering
    \includegraphics[width=\textwidth]{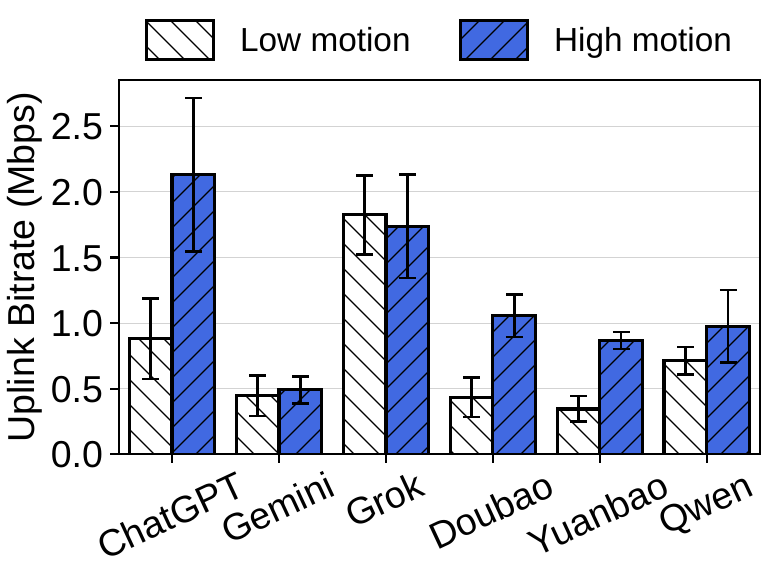}
    \caption{Constant or variable bitrates}
    \label{fig:video-coding-strategy}
  \end{minipage}
  \hfill
  \begin{minipage}[c]{0.31\textwidth}
    \centering
    \includegraphics[width=\textwidth]{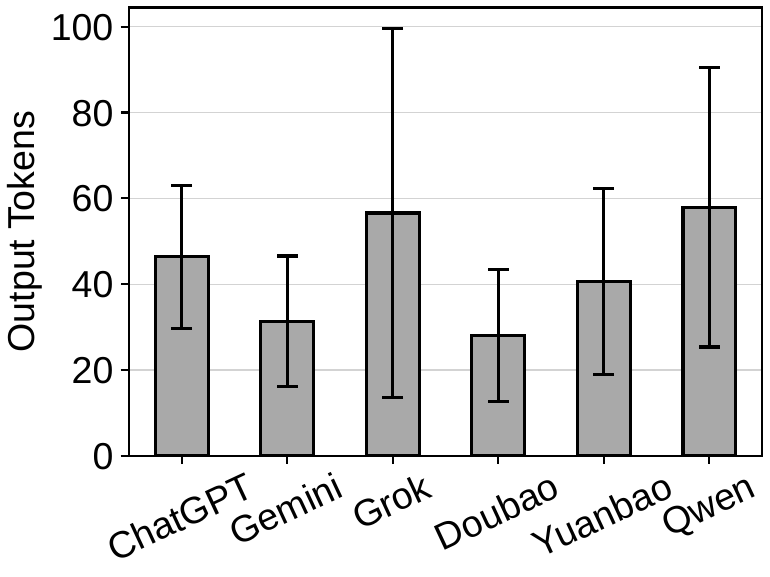}
    \caption{Output tokens}
    \label{fig:output-tokens}
  \end{minipage}
\end{figure*}

\parahead{Uplink \& Downlink rate.}
The uplink and downlink data rates are illustrated in Figure~\ref{fig:video-bitrate}.
It is apparent that the uplink bitrates differ substantially across various applications.
\chatgpt and \grok display the highest uplink bitrates at \textasciitilde1.9\,Mbps, while \gemini has the lowest at \textasciitilde0.5\,Mbps -- a 4$\times$ spread between the highest and lowest.
\doubao, \yuanbao, and \qwen sit in the middle (\textasciitilde0.8--0.9\,Mbps).
Despite having comparable average bitrates, the uplink bitrate of \chatgpt exhibits greater fluctuation than that of \grok (one possible reason is the difference in video coding strategy, which we discuss in the next paragraph).
The downlink rates are at least 5$\times$ lower than the uplink for every application, ranging from 27\,Kbps (\gemini) up to 164\,Kbps (\qwen).
This is expected, as the downlink stream carries only audio whereas the uplink includes both video and audio.

We next investigated the video coding strategy for each application, focusing on how encoding bitrates adapt to the complexity of the video content.
Our test involved two distinct video types: low-motion, featuring a static background, and high-motion, characterized by dynamic backgrounds and significant movement.
As illustrated in Figure~\ref{fig:video-coding-strategy}, \chatgpt, \doubao, and \yuanbao all employ an aggressive content-adaptive bitrate strategy: their high-motion uplink is roughly 2.4$\times$ the low-motion uplink, indicating an efficient encoder that conserves bandwidth on simple content.
\qwen also adapts, but more modestly (high-motion is \textasciitilde1.4$\times$ low-motion).
\grok and \gemini both lack content adaptation, but at very different operating points.
\grok holds a near-constant \emph{high} bitrate (\textasciitilde1.8\,Mbps; the high- and low-motion means differ by only \textasciitilde6\%), pointing to a less sophisticated coding strategy that wastes bandwidth on simple content.
\gemini, by contrast, holds a near-constant \emph{low} bitrate (\textasciitilde0.5\,Mbps; high-motion is only \textasciitilde10\% above low-motion), so its encoder does not push more bits into high-motion content even when the channel could carry them.
This behavior may suggest that \gemini implements a configured bitrate ceiling.

\begin{table*}[t]
    \centering
    \begin{threeparttable}
    \begin{tabular}{l|cccccc}
        \hline
        & \chatgpt & \gemini & \grok & \doubao & \yuanbao & \qwen \\ \hline
        Protocol           & RTP/RTCP & QUIC     & RTP/RTCP\textsuperscript{\dag} & RTP/RTCP\textsuperscript{\ddag} & RTP/RTCP & RTP/RTCP \\
        \# SSRCs (uplink)  & 2        & --       & 5        & 2--4     & 2        & 2 \\
        Avg bitrate (Kbps, up$|$down) & 1930$|$71 & 479$|$27 & 1820$|$43 & 866$|$107 & 759$|$73 & 895$|$164 \\
        Frame rate          & \textasciitilde30 & \textasciitilde1 & \textasciitilde1 & 6 & 10 & \textasciitilde30 \\
        Sending pattern    & Bursty   & Bursty   & Bursty   & Paced    & Bursty   & Bursty \\
        Input modality     & Audio    & Audio    & Text     & Text     & Text     & Text \\
        Avg output tokens  & 46       & 31       & 57       & 28       & 41       & 58   \\ \hline
    \end{tabular}
    \begin{tablenotes}
        \item[\dag] \grok's 5 SSRCs comprise audio + a baseline simulcast layer + two higher-resolution simulcast layers + RTX (standard WebRTC simulcast/RTX, RFC~8853 / RFC~4588).
        \item[\ddag] \doubao prepends a non-standard 2-byte \texttt{0xca 00} prefix before each RTP header.
    \end{tablenotes}
    \caption{Internal mechanism of the tested applications}
    \label{tab:mechanism}
    \end{threeparttable}
\end{table*}

\begin{figure*}[t]
  \centering
  %---- 三个小图水平排开 ----%
  \begin{minipage}[c]{0.31\textwidth}
    \centering
    \includegraphics[width=\textwidth]{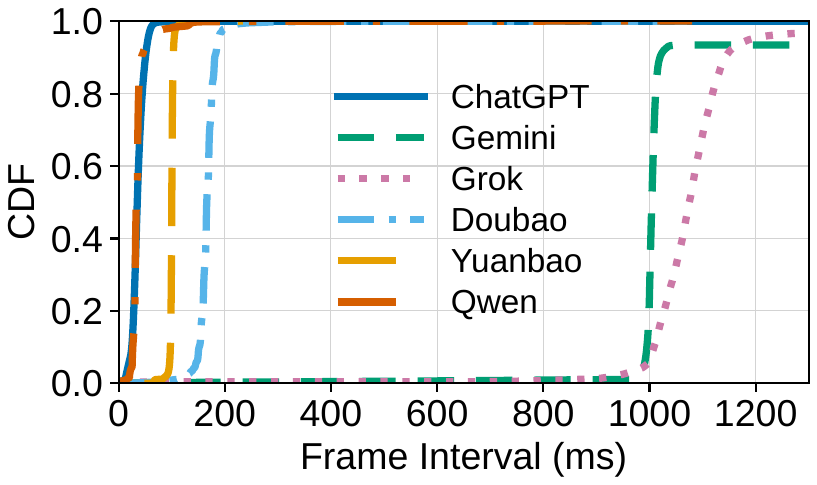}
    \caption{Frame intervals across apps}
    \label{fig:frame-rate}
  \end{minipage}
  \hfill
  \begin{minipage}[c]{0.32\textwidth}
    \centering
    \includegraphics[width=\textwidth]{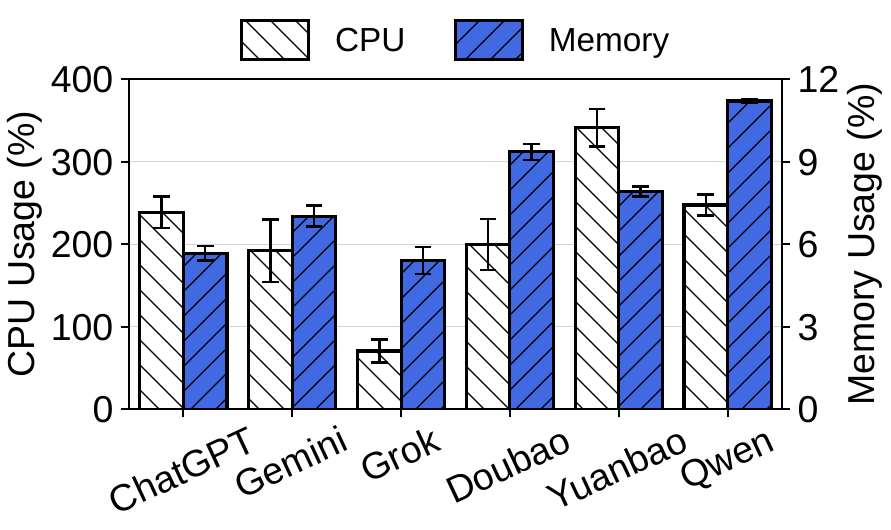}
    \caption{CPU and memory usage}
    \label{fig:cpu-overhead}
  \end{minipage}
  \hfill
  \begin{minipage}[c]{0.32\textwidth}
    \centering
    \includegraphics[width=\textwidth]{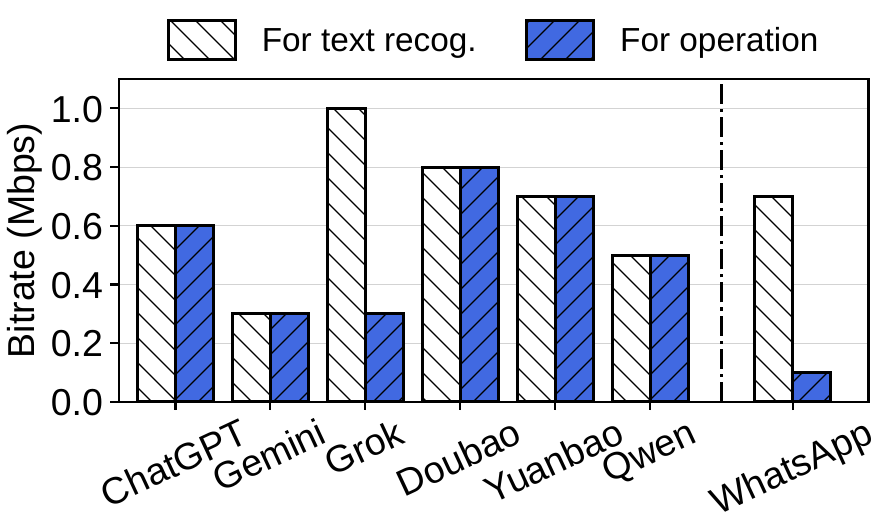}
    \caption{Minimal acceptable bitrate}
    \label{fig:quality-bitrate}
  \end{minipage}
  % \hfill
\end{figure*}

\parahead{Frame rate.}
To investigate if AI video chat deviates from the typical 15 (or more) fps standard of traditional video chat, we measured the frame rate of the video stream sent from the user to the AI.
For applications using RTP (\chatgpt, \grok, \doubao, \yuanbao, and \qwen), we identified individual frames directly via RTP timestamps (one timestamp per encoded frame).
For \gemini, which uses an opaque QUIC stream, we approximated the frame rate by calculating its packet burst frequency.
As shown in Figure~\ref{fig:frame-rate}, two applications run at the typical RTC frame rate: \chatgpt and \qwen both operate around \textasciitilde30\,fps. The remaining applications operate at substantially lower frame rates: \yuanbao (10\,fps), \doubao (6\,fps), \grok (\textasciitilde1\,fps), and \gemini (\textasciitilde1\,fps).
It is important to note that these measured network frame rates represent an upper bound; the actual rate of frames processed by the AI, which we cannot directly measure, may be even lower.
This trend toward lower frame rates can be explained by two primary factors: one a technical limitation, the other a design choice.
First, the computational demands of AI inference create a bottleneck, limiting the frequency at which video frames can be processed.
Second, unlike human vision, which requires high frame rates to perceive fluid motion, an AI's goal is analytical.
It can often achieve the same outcome by processing fewer, information-rich frames, making a high-rate video stream unnecessary.

% \begin{figure}[t]
%     \centering
%     \includegraphics[width=0.8\linewidth]
%     {figures/frame-intervals.pdf}
%     \caption{frame intervals CDF}
%     \label{fig:frame-rate}
% \end{figure}

\parahead{Video packet sending pattern.}
We also examined the video packet transmission patterns across different applications.
For the RTP-based systems, we measured the inter-packet arrival times within a single RTP timestamp (i.e., among the packets that carry one encoded video frame): when packets arrive within a few milliseconds of each other (\textless 1\,ms median across our samples) we label the sender \emph{bursty}; when packets are spread over a substantial fraction of the frame interval we label it \emph{paced}.
For \gemini, whose QUIC stream is encrypted and lacks visible RTP timestamps, we instead applied the same 10\,ms-bin packet-burst analysis we used for its frame-rate estimate; here a burst is a contiguous run of bins each carrying $\geq$60\,Kbps, and we declare the sender bursty when frames are delivered as tight clusters of bursts at the inferred frame rate.
With these measurements we find that \chatgpt, \grok, \yuanbao, \qwen, and \gemini all exhibit a \emph{bursty} pattern: every video frame is delivered as a tight burst, with all of its packets reaching the wire within roughly one millisecond.
\doubao is the only application that paces its packets, spreading each frame's packets across roughly 70\,ms within a 167\,ms (\textasciitilde6\,fps) frame interval.
Bursty transmission lets the encoder push frames out as soon as they are ready, but it stresses bottleneck queues and can cause short-lived congestion; paced transmission smooths out the queue occupancy at the cost of slightly higher per-frame latency.
The two strategies represent a well-known trade-off between link utilization and queueing-induced loss~\cite{sigcomm2025ace}.

\parahead{Input modality.}
An AI video chat's backend can process user input via one of two primary modalities: a cascaded pipeline (text + video frames), which first converts speech to text, or an end-to-end model (audio + video frames) that processes the raw audio signal directly.
The key distinction is the end-to-end model's ability to interpret non-verbal information that is lost during text transcription.
To determine which modality each application uses, we conducted an experiment using audio inputs that a speech-to-text system would ignore: sounds from musical instruments, natural sounds like rain and birds, and human emotional expressions like laughter and crying.
The results show that \chatgpt and \gemini both successfully identified and responded to these non-verbal sounds, indicating they employ an end-to-end, audio-native model.
In contrast, \grok, \doubao, \yuanbao, and \qwen failed to recognize these inputs.
This strongly suggests they rely on a cascaded pipeline that discards all non-speech information at the initial text transcription stage.

\parahead{Output tokens.}
\begin{table}[t]
    \centering
    \resizebox{\columnwidth}{!}{
    \begin{tabular}{c|cccccc}
        \hline
        & \chatgpt & \gemini & \grok & \doubao & \yuanbao & \qwen \\
        \hline
        2025-08 & 46.35 & 26.50 & 33.30 & 45.18 & 28.36 & -- \\
        2026-03 & --    & 31.36 & 56.58 & 28.03 & 40.62 & 57.94 \\
        $\Delta$   & --    & +4.86 & +23.28 & --17.15 & +12.26 & -- \\
        \hline
    \end{tabular}
    }
    \caption{Average output tokens per AI reply across two rounds}
    \label{tab:output-tokens-change}
\end{table}
The verbosity of an AI agent's reply is itself a user-facing characteristic: more tokens mean more time-on-screen for the user to read or longer audio for the user to listen to.
% We restrict the count to the chatbot-style tasks (\textit{real-time visual understanding}, \textit{contextual understanding} subset, and \textit{visual named entity recognition}) and exclude the audio-noisy categories, math, and proactive items.

\textit{Across applications.}
Figure~\ref{fig:output-tokens} shows the output tokens per AI reply across the six applications using the latest available results for each application.
% (the 2026-03 round for five apps; the 2025-08 round for \chatgpt, which was not retested in 2026-03)
The average reply length spans roughly $2\times$ across the six apps: \qwen and \grok produce the longest replies (\textasciitilde58 tokens on average), while \doubao yields the shortest (\textasciitilde28 tokens), with \chatgpt, \yuanbao, and \gemini in between (31--46 tokens).
This means a user chatting with \qwen or \grok hears or reads roughly twice as much content per turn as one chatting with \doubao for the same question, which directly translates into longer turn duration and therefore a noticeably slower conversational pace.

\textit{Across time.}
For the four applications tested in both rounds, output verbosity has shifted substantially (Table~\ref{tab:output-tokens-change}).
\grok grew the most dramatically (33.3 $\rightarrow$ 56.6 tokens, +23.3 or +70\%), followed by \yuanbao (+12.3) and \gemini (+4.9).
\doubao moved in the opposite direction, becoming \textasciitilde38\% more concise (45.2 $\rightarrow$ 28.0 tokens, --17.2).
These shifts indicate that output verbosity is not a fixed property of an application but a knob the vendors actively tune between deployments -- through prompt engineering or backend model swaps -- and one that directly affects user-perceived response time.

\subsection{System Overhead}
\label{sec:overhead-result}
We then analyzed the on-device resource utilization, focusing on CPU and memory consumption across the six applications.
All overhead measurements are taken on the local Redmi Note 10 Pro (Android 13, octa-core MediaTek Dimensity 1100, 8\,GB RAM) with the \texttt{top} command, which reports the per-process aggregate exposed by the Linux kernel: a single fully-busy core counts as 100\%, so a process that saturates four cores in parallel registers as 400\%, and the theoretical ceiling on this device is 800\%.
As illustrated in Figure~\ref{fig:cpu-overhead}, the gap between the highest and lowest CPU usage during a video chat exceeds $4.8\times$ (341\% vs 70\%), and the gap on memory usage exceeds $2\times$ (11.2\% vs 5.4\%).
\yuanbao exhibits the highest CPU usage at $\sim$341\% (the equivalent of 3.4 cores fully busy), with \qwen, \chatgpt, and \doubao all also above 190\%.
On memory, \qwen has the largest footprint at 11.2\%, followed by \doubao at 9.4\% and \yuanbao at 7.9\%.
\grok sits at the opposite end on both axes: its CPU usage stays around 70\% and its memory usage at 5.4\%, the lowest among all six applications, indicating a markedly lighter resource profile than its peers.

\subsection{Impact of Bandwidth Constraints}
\label{sec:low-rate-result}
In this section, we evaluate the degradation of application performance under bandwidth constraints. 
Specifically, our objectives are: 
(1) to quantify the extent to which network limitations impair the visual functionality of AI applications; 
(2) to conduct stress testing on the target applications to verify their operability under low-bandwidth conditions; 
and (3) to find out the performance differences between AI chatbots and traditional RTC applications -- for this purpose, WhatsApp video calls are incorporated as a comparative baseline.

To establish low-bandwidth network environments, the test smartphone was first connected to a dedicated Wi-Fi access point, with data transmission rates regulated through \texttt{tc}~\cite{tc_linux}. 
For each application, we gradually reduced the available bandwidth from its pre-measured average value to 100 kbps, using decrements of 100 kbps per iteration. 
To assess the visual capability of the tested applications, AI chatbots were tasked with recognizing the title of a paper (12-point font) printed on an A4 sheet, where the distance between the smartphone’s camera and the paper was fixed at 30 cm. 
For WhatsApp video calls, two participants were asked to visually identify the same paper title, with the metric being the minimum bitrate at which the title remained distinguishable to both.

% These thresholds were based on the average uplink bitrate data for each application, as shown in Figure~\ref{fig:video-bitrate}.

As shown in Figure~\ref{fig:quality-bitrate}, the contrast between AI chatbots and the traditional video-call baseline is striking.
For WhatsApp, the two participants need 700 kbps to distinguish the printed title, yet the video call itself stays connected even at the minimum 100 kbps -- a $7\times$ gap between the visual-recognition threshold and the operating threshold.
For five of the six AI chatbots (\chatgpt, \gemini, \doubao, \yuanbao, \qwen) no such gap exists: the minimum bandwidth at which the agent can read the title is exactly the threshold below which the agent stops working at all.
This minimum varies substantially across these five, from 300 kbps (\gemini, the most bandwidth-efficient) and 500 kbps (\qwen) up to 800 kbps (\doubao, the most demanding); \chatgpt and \yuanbao sit in the middle at 600--700 kbps.
\grok is the lone exception and behaves more like the traditional baseline: it stays operational down to 300 kbps but needs 1000 kbps -- the highest threshold of any AI chatbot in our test -- to recognize the title.
Notably, this is a change from the previous round, when \grok could not read the title at any bitrate we tested; it has since gained the capability, but at the cost of being the most bandwidth-hungry on this task.

In conclusion, our results highlight two findings:
\begin{itemize}
    \item Traditional RTC applications can function reliably at very low bitrates (around 100 kbps), whereas AI chatbots cannot -- likely due to inherent limitations in their application design that couple visual processing to overall connection quality.
    \item Within their feasible bandwidth range, most AI chatbots can recognize printed text at lower bitrates than human participants (300--700 kbps versus 700 kbps for WhatsApp users); the exceptions are \doubao and \grok, both of which need more bandwidth than humans for the same recognition task.
\end{itemize}

% \subsection{Miscellaneous Results}

% We also monitor CPU usage on the local Android device when using different AI video chat applications. 

% We report CPU usage in absolute numbers, e.g., 200\% implies
% full utilization of two cores. 
% We also compare it with traditional video chat apps like WeChat and WhatsAPP. 

% \begin{figure}[t]
%     \centering
%     \includegraphics[width=1\linewidth]{figures/setup-time.jpg}
%     \caption{CPU usage}
%     \label{fig:cpu-overhead}
% \end{figure}

%-------------------------------------------------------------------------------
\section{User Study}
%-------------------------------------------------------------------------------
\label{sec:user-study}

\begin{figure*}[t]
    \centering
    \includegraphics[width=0.85\linewidth]{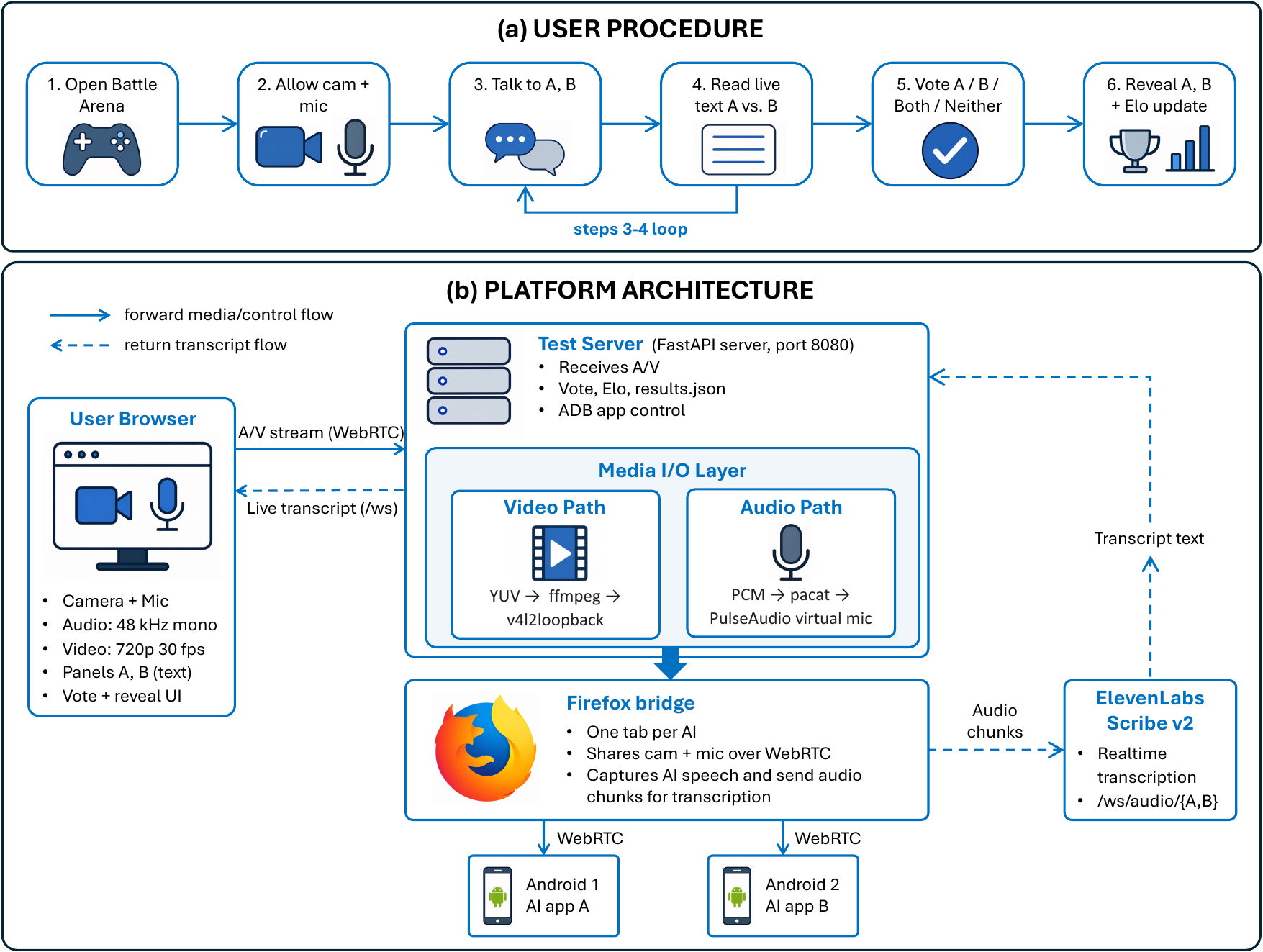}
    \caption{Overview of the arena platform.}
    \label{fig:platform}
\end{figure*}

The objective measurements in \S\ref{sec:results} -- response quality on our constructed dataset, end-to-end latency, system overhead -- characterize what each application can do under controlled conditions, but they do not fully predict how a real user will feel about a given app.
Two factors break the mapping between objective metrics and perceived experience.
First, our quality benchmark is built on a curated dataset queried in English, while real users hold conversations in their own language with their own questions and visual contexts; an app whose text style or language coverage is well-matched to a particular user population may rank higher in real use than its dataset score predicts.
Second, the rating is a holistic preference judgment that aggregates response accuracy, fluency, conciseness, and tone, while the dataset benchmark scores each task in isolation -- an app that scores lower on individual tasks may still produce text that users prefer overall.
To quantify how these factors aggregate into user preference, we built an online arena where users hold blind side-by-side video chats with two randomly selected AI assistants and rate which one they preferred (\S\ref{sec:battle-platform}).
We then analyze the resulting ratings to (i) produce a user-driven ranking of the apps, and (ii) test whether response latency -- the headline metric we measured in \S\ref{sec:response-delay-result} -- actually drives user preference (\S\ref{sec:rating-results}).

\subsection{Arena Platform Design}
\label{sec:battle-platform}

Figure~\ref{fig:platform} gives an overview of the platform from the user's perspective and the architecture that supports it.
Because \chatgpt cannot run on Android emulators (\S\ref{sec:testbed-design}), the arena features the remaining five applications.

\parahead{User procedure.}
A user visits the public URL and grants camera and microphone permissions; their preview appears on the page.
After clicking \emph{Start}, the platform pairs them blind with two of the five AI applications (sides A and B), launches the corresponding apps on two Android emulators, and opens the user's audio/video to both.
The two AIs respond simultaneously; we transcribe each side's response audio in real time and stream the transcripts back to the user as side-by-side text bubbles.
Each battle is hard-capped at three minutes.
After the user ends the session or the cap is reached, they pick which side was better -- \emph{A is Better}, \emph{B is Better}, \emph{Both Good}, or \emph{Neither Good} -- and the AI identities are revealed.
Per-pair match counts are tracked, so the platform schedules the least-played pair on each subsequent session, keeping pair coverage roughly balanced.

\parahead{Architecture.}
The core challenge in serving real users is that the AI applications run on Android emulators in our cloud testbed, while the user is on a regular browser.
The platform bridges these two ends with three pipes.
(i) The user's webcam and microphone are streamed to the backend via WebRTC, then injected into the emulators' virtual camera (\texttt{v4l2loopback}) and virtual microphone (\texttt{PulseAudio}) so each AI app perceives them as the device's hardware.
(ii) The AI apps' audible replies, captured on each emulator's audio output, are transcribed in real time by a streaming speech-to-text service (\texttt{ElevenLabs Scribe v2 Realtime}) and pushed back to the user as text bubbles.
(iii) Ratings are persisted to a leaderboard whose ELO ranking is recomputed from scratch on every read.
We deploy two backend instances per host, each owning its own pair of Android emulators; an nginx upstream pool with HTTP-503-aware failover routes each user to whichever instance is currently free, so two users can hold concurrent battles on the same host.

\parahead{Delay injection.}
A naive sign-test or correlation analysis of user ratings against natural latency would be confounded by app identity: an app that happens to be slow may also produce better text, and observational data alone cannot separate the two.
To estimate the \emph{causal} effect of latency on user preference, we randomly inject an artificial delay on every battle.
Each battle independently draws an injection magnitude $\delta \in \{0, 1, 2\}$~s and an injection side $s \in \{A, B\}$; whenever the side $s$ produces a transcript, the platform delays its delivery to the user by $\delta$ before rendering.
Because the assignment is randomized, the effect of $\delta$ on the rating is identified independently of which app happens to be on which side.

\subsection{Results}
\label{sec:rating-results}

\begin{table}[t]
    \centering
    \begin{tabular}{cl|ccc}
        \hline
        \# & App & \textbf{ELO score} & Win rate & \# Matches \\ \hline
        1 & \doubao  & \textbf{1549} & 68.5\% & 84 \\
        2 & \gemini  & \textbf{1510} & 51.2\% & 86 \\
        3 & \qwen    & \textbf{1506} & 47.1\% & 86 \\
        4 & \grok    & \textbf{1496} & 37.9\% & 87 \\
        5 & \yuanbao & \textbf{1438} & 35.3\% & 85 \\ \hline
    \end{tabular}
    \caption{Arena leaderboard (as of 2026-04-29).}
    \label{tab:leaderboard}
\end{table}

\begin{table}[t]
    \centering
    \begin{tabular}{l|cc}
        \hline
        Effect & $\beta$ (log-odds) & 95\% CI \\
        \hline
        Injected delay (per second)     & $-0.065$ & $[-0.379, +0.250]$ \\
        \hline
        \multicolumn{3}{l}{\textit{App fixed effects (vs.\ \doubao reference)}} \\
        \hline
        \quad \gemini  & $-0.80$ & $[-1.64, +0.03]$ \\
        \quad \grok    & $-1.07$ & $[-1.92, -0.23]$ \\
        \quad \qwen    & $-0.69$ & $[-1.52, +0.14]$ \\
        \quad \yuanbao & $-1.15$ & $[-1.98, -0.33]$ \\
        \hline
    \end{tabular}
    \caption{Logistic regression of $A\_\text{wins} \sim \text{injected delay} + \text{app fixed effects}$ ($n=111$ decisive battles). The injected-delay coefficient is small and not distinguishable from zero, whereas app identity is an order of magnitude larger and significant for three of four non-reference apps.}
    \label{tab:latency-effect}
\end{table}

We report results from the arena built up through 2026-04-29.
The leaderboard in Table~\ref{tab:leaderboard} aggregates 214 rated battles across the five applications; the latency analysis uses 124 battles since the injection experiment was instrumented (111 decisive ``A''/``B'' votes plus 13 ties).

\parahead{Result-1: User preference largely tracks objective quality.}
The user-driven ranking from the arena (\doubao $>$ \gemini $>$ \qwen $>$ \grok $>$ \yuanbao, Table~\ref{tab:leaderboard}) is broadly consistent with the objective response-quality ranking on our constructed dataset (\gemini $>$ \doubao $>$ \qwen $>$ \grok $>$ \yuanbao, Table~\ref{tab:scores}): three of five positions (\qwen 3rd, \grok 4th, \yuanbao 5th) match exactly, and the only inversion is at the top.
We refrain from reporting a numerical rank correlation here because, with only five apps, a single rank swap moves the statistic substantially and the arena's ELO ranking continues to fluctuate as more battles are recorded; we instead report the observation directly.
The inversion concerns the top pair: \doubao edges out \gemini in user preference (1549 vs.\ 1510 ELO; 68.5\% vs.\ 51.2\% win rate) despite \gemini scoring higher on objective text-quality benchmarks.
We attribute this to two factors that our dataset benchmark does not capture.
First, a non-trivial share of arena participants conversed in Chinese, broadening the linguistic coverage beyond our English-only dataset; \doubao's response style appears to be better matched to that population.
Second, user preference aggregates response style, conciseness, and contextual relevance into a single judgment, while the dataset benchmark scores per-task accuracy in isolation -- an app that does not lead on individual tasks may still produce text that users prefer overall.
The arena thus complements, rather than replaces, the text-only quality metrics.

\parahead{Result-2: Latency does not drive user preference within the observed range.}
Of the 111 decisive battles, 72 received a nonzero injection (the remaining 39 formed the 0-s control arm).
Among those 72, the artificially slowed AI won 54.2\% of the time (95\% Wilson CI $[42.7\%, 65.2\%]$); the interval comfortably contains 50\%, indicating no significant effect of injected latency on which side the user preferred.
To make this estimate independent of which app happened to be slowed, we fit a logistic regression of $A\_\text{wins}$ on injected delay (signed in seconds, positive when B is slowed) with app fixed effects on all 111 decisive battles (the 39 control battles contribute $\delta = 0$ to the regression); the coefficient on injected delay is $\beta = -0.065$ log-odds per second (95\% CI $[-0.379, +0.250]$), corresponding to an odds ratio of $0.94$ per second of injected delay (95\% CI $[0.69, 1.28]$).
The app fixed effects are an order of magnitude larger and statistically significant for three of the four non-reference apps (Table~\ref{tab:latency-effect}), confirming that user preference is driven by which app the user is talking to, not by how fast it replies.
% As an auxiliary check, we examined whether the rate of tie ratings (\emph{Both Good} or \emph{Neither Good}) trended upward with injected delay (one would expect such a trend if forced delay degraded user satisfaction).
% We observe a mild upward drift -- $9.3\%$, $9.5\%$, and $12.8\%$ at $0$, $1$, and $2$\,s of injection respectively -- but the differences are small relative to the per-cell sample sizes ($n = 39$--$43$) and the 95\% Wilson intervals overlap heavily, so we cannot read this as evidence either for or against a causal effect.

This null result is consistent with the latency profile we measured in \S\ref{sec:response-delay-result}: end-to-end response delays for current AI video chat applications already sit in the multi-second range and are dominated by AI-side processing rather than network transit.
Within that operating range, an additional second of network delay does not measurably change which app the user prefers.
The practical implication is that latency optimization at the network layer alone is unlikely to translate into noticeable user-experience gains; reducing AI-side processing time, or improving response \emph{quality}, has a higher payoff per engineering hour.

%-------------------------------------------------------------------------------

\section{Related Work}
%-------------------------------------------------------------------------------

\parahead{Measurements over LLM/MLLM serving.}
Large Language Models (LLMs) and Multimodal Large Language Models (MLLMs) are typically deployed on cloud servers, and extensive research has focused on optimizing their reference time~\cite{li2024llm,wang2024towards,lin2024qserve,ding2024sustainable,qiu2024efficient}.
Numerous studies~\cite{peng2024survey,li2023beyond,zhou2023don,white2024livebench} have reported the performance of LLM/MLLM inference, with a primary focus on metrics defined from the LLM serving perspective -- specifically Time To First Token (TTFT) and Time Per Output Token (TPOT)~\cite{horton2024kv}.

However, these studies primarily target the internal processing efficiency of the model itself, with metrics that focus on the server-side inference process rather than the end-to-end experiences in human-agent interaction.
% Notably, the AI response delay measured in this paper differs from TTFT and TPOT. 
Existing optimizations for TTFT/TPOT are tailored to server-side LLM performance, and their inapplicability to the human-AI agent interaction paradigm highlights the gap in current research on AI video chat -- indicating the need for more targeted approaches for this new scenario.

% LLM/MLLM are deployed on server cloud. 
% Technologies they use to accelerate reference time. 
% TTFT, TPOT. These metrics are from an LLM serving perspective, which are different from the AI response delay we measured in this work. 
% The optimizations are different, which indicates a more specific approach for this new paradigm should be proposed. 

%-------------------------------------------------------------------------------
\section{Limitations}
%-------------------------------------------------------------------------------

% Since these applications are like black box and we are from an outsider view, we can not know how it exactly works inside. 
% We can not know the exact value of each component of response delay, including network delay and AI inference delay. 
% frame rate in network doesn't mean the frequency the AI takes video frame input. 

\parahead{Blackbox nature of AI video chat systems.}
A key limitation of this study is the blackbox nature of the tested AI video chat apps -- external researchers cannot access their internal details (e.g., multimodal data processing workflows, LLM inference optimization). 
This opacity makes it impossible to break down "AI response delay" into core components and verify if observed network frame rate matches the system’s actual video frame processing frequency.

% \parahead{Lack of subjective metrics.}
% In this paper, we do not involve scores from the subjective experiments from users. 
% This is because the subjective user experience might also be affected by how users use the AI video chatbot. 
% Alternatively, we introduce the human score in~\S\ref{sec:quality-result} to present the score.

\parahead{Client variability.}
This study’s results may be limited by client-side variability, as it cannot fully account for differences in end-users’ device configurations, OS versions, or software states. 
The local testbed only used one rooted Android device with fixed WiFi and Bluetooth, while real users have diverse devices such as iOS phones.

\section{Conclusion}
%-------------------------------------------------------------------------------

This paper is motivated by the critical need to conduct the first systematic performance measurement of AI video chat systems, thereby providing the research community with a baseline understanding of their real-world performance. 
In a nutshell, from our benchmark, \yuanbao shows the fastest response yet lowest quality, while \chatgpt, \gemini, and \doubao all behave relatively well on the quality, though still far behind human performance. 
We further identify their unique bottlenecks and establish a foundation for future optimization efforts. 

% \begin{acks}
% \end{acks}

\clearpage

\bibliographystyle{ACM-Reference-Format}
\bibliography{reference}

\clearpage

\appendix









\section{Ethics}
This work does not raise any ethical issues. On our arena platform, we do not record, store, or reuse user video or audio content. Only anonymous battle metadata (which apps the user compared, user rating, response latency) is kept for the leaderboard.

\end{document}